\documentclass{jkas}

\def\year{2014} 
\def\month{April} 
\def\volume{47} 
\def\issue{2} 
\def\beginpage{49} 
\def\endpage{67} 
\def\received{January 16, 2014} 
\def\accepted{February 25, 2014} 
\date{Received \received ; accepted \accepted}

\newcommand{\fn}[2]{\mathinner{#1\mathopen{\left(#2\right)}}}
\newcommand{\Mpc}{\mathinner{\mathrm{Mpc}}}
\newcommand{\Mpch}{\mathinner{h^{-1}\,\mathrm{Mpc}}}
\newcommand{\kpch}{\mathinner{h^{-1}\,\mathrm{kpc}}}

\newcommand{\Myr}{\mathinner{\mathrm{Myr}}}
\newcommand{\mK}{\mathinner{\mathrm{mK}}}
\newcommand{\MHz}{\mathinner{\mathrm{MHz}}}
\newcommand{\Kelvin}{\mathinner{\mathrm{K}}}
\newcommand{\Lya}{Ly$\alpha$ }
\newcommand{\dtb}{\delta T_b}
\newcommand{\mdtb}{\langle \dtb \rangle}
\newcommand{\rdtb}{\delta T_{b,\mathrm{rms}}}
\newcommand{\dtbth}{\delta T_{b,\mathrm{th}}}
\newcommand{\dtbn}{\delta T_{b,n}}
\newcommand{\kmsMpc}{\mathinner{\mathrm{km}\,\mathrm{s}^{-1}\,\mathrm{Mpc}^{-1}}}
\newcommand{\fiducial}[3]{\mathinner{\left( \frac{#1}{#2} \right)^{#3}}}

\def\arcmin{\rm arcmin}
\def\etal{\rm et al.}
\def\apj{ApJ}
\def\mnras{MNRAS}
\def\apjs{ApJS}
\def\aj{AJ}
\def\apss{ApSS}
\def\baas{BAAS}
\def\apjl{ApJL}
\def\prd{Phys. Rev. D}
\def\prl{Phys. Rev. Lett.}

\usepackage{flushend} 

\title {
2D Genus Topology of 21-cm Differential Brightness Temperature During Cosmic Reionization
}

\author[a]{Sungwook E. Hong}
\author[b]{Kyungjin Ahn}
\author[a]{Changbom Park}
\author[c]{Juhan Kim}
\author[d]{Ilian T. Iliev}
\author[e]{Garrelt Mellema}

\affil[a]{Department of Physics, Korea Institute for Advanced Study, Seoul 130-722, Korea; swhong@kias.re.kr and cbp@kias.re.kr}
\affil[b]{Department of Earth Science Education, Chosun University, Gwangju 501-759, Korea; kjahn@chosun.ac.kr}
\affil[c]{Center for Advanced Computation, Korea Institute for Advanced Study, Seoul 130-722, Korea; kjhan@kias.re.kr}
\affil[d]{Astronomy Centre, Department of Physics \& Astronomy, Pevensy II bldg, University of Sussex, Falmer, Brighton BN1 9QH, UK; I.T.Iliev@sussex.ac.uk  }
\affil[e]{Department of Astronomy \& Oskar Klein Centre, AlbaNova, Stockholm University, SE-106 91 Stockholm, Sweden; garrelt@astro.su.se}

\def\corrauthor{
K. Ahn
}

\def\runningauthor{
S. E. Hong et al. (2014)
}

\def\runningtitle{
2D Genus Topology of the 21-cm Differential Brightness Temperature
}

\def\keywords{
cosmology: theory --- cosmology: observations --- intergalactic medium --- large-scale structure of Universe
-- methods: statistical --- radiative transfer
}

\def\abstracttext{
A novel method to characterize the topology of the early-universe
intergalactic medium during the epoch of cosmic reionization is
presented. The 21-cm radiation background from high redshift is
analyzed through the calculation of the 2-dimensional (2D) genus.
The radiative transfer of hydrogen-ionizing photons and
ionization-rate equations  are calculated in a suite of numerical
simulations under various input parameters.
The
2D genus is calculated from the mock 21-cm images of the high-redshift Universe.
We construct the 2D genus curve by varying the threshold
differential brightness temperature, and compare this to the 2D genus
curve of the underlying density field. We find that
(1) the 2D genus curve
reflects the evolutionary track of cosmic reionization and (2) the 2D
genus curve can discriminate between certain reionization scenarios and
thus indirectly probe the properties of radiation-sources.
Choosing the right beam shape of a radio antenna is crucial for this analysis.
To this end, the Square Kilometre Array (SKA) is found to be a suitable apparatus for
this analysis in terms of sensitivity, even though some
deterioration of the data for this
purpose is unavoidable under the planned size of the antenna core.
}

\begin{document}
\jkashead


\section{Introduction}

The epoch of cosmic reionization (EOR) commences with the birth of the first astrophysical, nonlinear objects such as the first stars and miniquasars.
These sources of radiation  create individual H\,{sc ii} bubbles by emitting hydrogen-ionizing photons.
This is then followed by the end of cosmic reionization,
when all percolating individual H\,{sc ii} bubbles fully merge with one another
and almost all hydrogen atoms in the Universe become ionized.

Direct observations of the EOR have not been made yet,
even though several indirect observations imply that
(1) the end of reionization occurred at a redshift $z \simeq 6.5$ \citep[e.g.][]{Fan:2005es},
(2) the intergalactic medium remained at high temperature during and after the EOR \citep{Hui:2003hn}, and
(3) this epoch started no later than $z \simeq 11$ \citep{Komatsu:2008hk}.
Theories suggest that the growth of H\,{sc ii} bubbles proceeded inhomogeneously (in a patchy way) due to the clustered distribution of radiation sources,
and the global ionized fraction $\langle x \rangle$ increased monotonically in time
\citep[e.g.][]{Iliev:2006sw,McQuinn:2007,Shin:2008}.
A few models predict, however, non-monotonic
increase in the global ionized fraction $\langle x \rangle$
due to the possible recombination after regulated star formation,
followed by emergence of stars of higher tolerance to photoheating
\citep[e.g.][]{Cen:2003ey,Wyithe:2003rr}.

The observation of the redshifted 21-cm line from neutral hydrogen is one of the most promising methods for the direct detection of the EOR,
and of the Cosmic Dark Ages that precedes this epoch as well.
Both temporal and spatial fluctuations in the 21-cm signal are believed to be strong in general, thus easy to detect,
during the EOR---due to the inhomogeneous growth of H\,{sc ii} bubbles and
relatively weak foregrounds at high frequencies. The strongest
signal will come from the late dark ages, or the very early EOR, when the spin
temperature will be much lower than the cosmic microwave background
radiation (CMB) due to the Ly-$\alpha$ coupling to a yet unheated
intergalactic medium (IGM), even though stronger foregrounds at lower
frequencies will be an obstacle to observations \citep[e.g.][]{Pritchard:2008}.
There are several large radio interferometer arrays
which aim at detecting the 21-cm signal from the EOR.
These projects include
the Giant Meterwave Radio Telescope\footnote{http://www.gmrt.ncra.tifr.res.in} (GMRT),
Murchison Widefield Array\footnote{http://mwatelescope.org} (MWA),
the Precision Array for Probing the Epoch of
Reionization\footnote{http://astro.berkeley.edu/$\sim$dbacker/eor/} (PAPER:
\citealt{Parsons:2010}),
the 21 Centimeter Array\footnote{http://web.phys.cmu.edu/$\sim$past/} (21CMA; formerly known as PaST),
the LOw Frequency ARray\footnote{http://www.lofar.org} (LOFAR),
and the Square Kilometre Array\footnote{http://www.skatelescope.org} (SKA; for the EOR detection strategy by SKA see \citealt{Mellema:2013}).

The lack of direct observations of the EOR results in poor constraints on the history of cosmic reionization.
Theoretical predictions for the history of cosmic reionization are made by semi-analytical calculations or full numerical simulations.
These studies select a variety of input parameters,
among which the most important one describes the properties of the sources of radiation.
The mock 21-cm data produced by such studies will be compared to future observations to constrain,
for example, the emissivity of high-redshift sources of radiation.

The patchy, 3D 21-cm radiation background can be analyzed
in various ways, such as with the 3D power spectrum, the 2D power spectrum,
the distribution of H\,{sc ii} bubble size, the cross-correlation of ionized
fraction and overdensity, etc \citep[e.g.][]{Iliev:2006sw, Zahn:2006sg, Iliev:2013}.
Each analysis method adds to the capability to discriminate between different
reionization scenarios.
The different methods are usually complementary to each other, allowing one to
understand the underlying physics in more detail.
Therefore, for data analysis it is always favorable to have as many different tools as possible.

In this paper, we characterize
the geometrical properties of the
distribution of neutral IGM and H\,{sc ii} bubbles by using the
topology of the 21-cm differential brightness temperature field. For this
purpose we measure the 2D genus of a series of snapshots of the
high-redshift Universe that are predicted in different models of
reionization.
By varying the
threshold value for the differential brightness temperature, we
construct 2D genus curves at different redshifts under different
reionization models. These models are simulated by a self-consistent
calculation of the radiative transfer and rate equations over our
simulation box.

Recently, similar methods for studying the topology
  of the high-redshift IGM have been suggested by
  \citet{Gleser:2006}, \citet{Lee:2008}, and \citet{Friedrich:2010}.
They calculated the 3D genus using either the neutral \citep{Gleser:2006,Lee:2008}
or ionized \citep{Friedrich:2010} fraction of IGM.
These studies show that the 3D genus of the underlying
  neutral (or ionized) fraction reflects the evolutionary stages of
  cosmic reionization. The 3D genus is also found to be useful in discriminating
  between different reionization scenarios.
Despite the similarity of our work to these studies, there are several
fundamental differences.
First, we calculate the 2D genus instead of the 3D genus. A 2D sky
analysis of a given redshift Universe, which is possible in the 21cm
observations, is beneficial especially when reionization proceeds
rapidly, because then a 3D analysis will suffer from the light-cone effect
\citep[e.g.][]{Datta:2012,LaPlante:2013} by mixing
different redshift information along the line of sight.
Second, we use an observable quantity,
  the 21-cm differential brightness temperature, which is more
  directly applicable to real data observed by future radio
  antennae, when calculating the 2D genus.
Third, we explicitly seek for the possibility of applying our
  method to the data to be observed by SKA, with an appropriate choice
  on the filtering scales for the 21-cm differential brightness temperature.

We employ the 2D genus method
developed by \citet{Melott:1989wv}.
The genus topology method was originally designed to test mainly
the Gaussian random-phase nature of the primordial density field
in 3D \citep{Gott:1986uz, Hamilton:1986, Gott:1987}
or in 2D \citep{Melott:1989wv}.
The 2D case has been studied for a variety of cosmological data sets:
on redshift slices \citep{Park:1991wp, Colley:1996gh, Hoyle:2001xx},
on sky maps \citep{Gott:1992, Park:2000ne},
on the CMB \citep{Coles:1989su, Gott:1989yj, Smoot:1992td, Kogut:1993,
  Kogut:1996uu, Colley:1996vj, Park:1997ad, Park:2001nq}, and on the
HI gas distribution in galaxy disks \citep{Kim:2007, Chepurnov:2008}.

Because the 2D genus is strongly affected by the number and distribution
of H II bubbles as well as the distribution of the underlying density
field, our method may provide insights about how reionization
proceeded and was affected by the properties of sources. We also seek for
the possibility of applying our method to observed data from SKA,
which so far has the highest sensitivity proposed.

The layout of our paper is as follows.
In Section~\ref{sec:sim} we describe our numerical simulations and the basic
procedure to calculate the 21-cm radiation background.
In Section~\ref{sec:genus} we describe how the 2D genus is calculated.
In Section~\ref{sec:res}, our simulation results are analyzed
and the possibility of using SKA observation results are
investigated.
We summarize our result and discuss future prospects in Section~\ref{sec:concl}.

\section{Simulations of redshifted 21-\lowercase{cm} Signals}\label{sec:sim}

\begin{table*}[!t]
\caption{Simulation parameters \label{tab:model}}
\centering
\begin{tabular}{cccccccccccc}
\toprule
Case &
Model &
$L_{\rm box}$ &
$N$-body &
RT &
$M_{\rm min}$ &
$f_{\gamma,{\rm high}}$ &
$f_{\gamma,{\rm low}}$ &
$\Delta t$ &
$\tau$ &
$z_{50\%}$ &
$z_{1\%}$ \\
 &
 &
(Mpc/$h$) &
mesh &
mesh &
($M_\odot$) &
 &
 &
($\Myr$) &
 &
 & \\
\midrule
1 & f250 & 64 & $512^3$ & $128^3$ & $2\times 10^9$ & 250 & - & 20 & 0.052 & 9.7 & 11.5 \\
 & (g125) &  & & & &  & & & &  & \\
2 & f5 & 64 & $512^3$ & $128^3$ & $2\times 10^9$ & 5 & - & 20 & 0.078 & 7.1 & 9.7 \\
 & (g2.5) &  & & & &  &  & &  & & \\
3 & f125\_1000S & 66 & $2048^3$ & $256^3$ & $10^8$ & 125 & 1000 & 10 & 0.111 & 12.0 & 17.3 \\
 & (g125\_1000S) &  & & & &  &  & &  & & \\
4 & f125\_125S & 66 & $2048^3$ & $256^3$ & $10^8$ & 125 & 125 & 10 & 0.104 & 11.6 & 15.2 \\
 & (g125\_125S) &  & & & &  &  & &  & & \\
\bottomrule
\end{tabular}
\tabnote{Cases 1 and 2 have only high-mass sources. Cases 3 and 4 have both high-mass and low-mass sources. Numbers in the ``Model'' are $f_\gamma$'s for high-mass halos, and if any follows, those are $f_\gamma$'s for low-mass halos (with ``S'' representing ``self-regulated''). The corresponding $g_\gamma$'s are also listed. The radiative transfer is calculated on the number of meshes listed under ``RT mesh.'' Note that even though Case 1 has the same $g_\gamma$ as Cases 3 and 4, its minimum mass $M_{\rm min}$ does not reach the low-end ($10^9 \,M_\odot$) of the other two, thus making its reionization history end much later.}
\end{table*}

\subsection{$N$-body Simulations}\label{sec:sim_nbody}

We completed a $2048^3$ particle simulation in a concordance $\Lambda$CDM model with WMAP 5-year parameters \citep{Dunkley:2008ie};
$\Omega_m = 0.258$,
$\Omega_\Lambda = 0.742$,
$\Omega_b = 0.044$,
$n_s = 0.96$,
$\sigma_8 = 0.79$, and $h = 0.719$,
where
$\Omega_m, \Omega_\Lambda, \Omega_b$ are the density parameters due to matter, cosmological constant, and baryon, respectively.
Here, $n_s$ is the slope of the Harrison-Zeldovich power spectrum,
and $\sigma_8$ is the root-mean-square (rms) fluctuation of the density
field smoothed at $8\Mpch$ scale.
The cubic box size is $66 \Mpch$ in a side length.
The simulation uses GOTPM, a hybrid PM$+$Tree $N$-body code \citep{Dubinski:2003fq,Kim:2008kf}.
The initial perturbation is generated with a random Gaussian distribution on a $2048^3$ mesh at $z_i = 500$.
The force resolution scale is set to $f_\epsilon = 3.2 \kpch$ in comoving scale.
A total of 12,000 snapshots, uniformly spaced in the
scale factor, are created.
The time step is predetermined so that the maximum particle
displacement in each time step is less than the force
resolution scale, $f_\epsilon$.

We then extract friend-of-friend (FoF) halos at
84 epochs: 24 epochs with a five million year interval from $z = 40$ to
$20$, and $60$ epochs with a ten million year separation between $z = 20$ and $z = 7$.
The connection length is set to 0.2 times the mean particle separation to identify cosmological halos.
Using this method, we find all halos with mass above $10^8 M_\odot$ (corresponding to 30 particles).
At each epoch, we calculate matter density fields on a $2048^3$ mesh
using the Triangular-Shaped-Cloud (TSC) scheme
\citep{Efstathiou:1985re}.
As the radiative transfer through the IGM does not need to be run at
too high resolution,
we further bin down the original density fields to $256^3$ after
subtracting the contribution of collapsed dark halos.
We transform this dark matter density into baryonic density by assuming
that in the IGM baryons follow the dark matter with the mean cosmic
abundance.
From the list of collapsed halos, we form a ``source catalogue'' by
recording the total mass of low-mass ($M < 10^9 M_\odot$) and high-mass
($M > 10^9 M_\odot$) halos in those radiative transfer cells containing halos.

We also ran a similar $N$-body simulation with the same cosmological parameters but a different configuration of $512^3$ particles on a $512^3$ mesh on a $64 \Mpch$ box using GOTPM,
resulting in somewhat poorer halo-mass resolution ($2 \times 10^9 M_\odot$).
Similarly, we created a binned-down density field on a $128^3$ mesh,
on which the radiative transfer is calculated.

We use results from the high resolution ($2048^3$ $N$-body) run for two cosmic reionization simulations (Case 3 and 4),
and results from the low resolution ($512^3$ $N$-body) run for the other
two (Case 1 and 2). For each cosmic reionization simulation, we
choose a fixed time step $\Delta t$.
These parameters are listed in Table~\ref{tab:model}.

\subsection{Reionization Simulations}\label{sec:sim_reion}

We performed a suite of cosmic reionization simulations based upon the
density field of the IGM and the source catalogue compiled from the N-body
simulations, described in Section~\ref{sec:sim_nbody}.
We used C$^2$-Ray \citep{Mellema:2005ht} to calculate the radiative transfer of H-ionizing photons from each source of radiation to all the points in the simulation box
and the change of ionization rate at each point, simultaneously.
The spatial resolution of the binned-down density field is the actual radiative-transfer resolution,
which is described in Table~\ref{tab:model} for each case.
At each $N$-body simulation step, we produce the corresponding 3D map of ionized fractions.

The history of cosmic reionization may depend on various physical properties of the radiative source,
such as the initial mass function (IMF) and the inclusion of Population III stars \citep[e.g.][]{Ahn:2012} or quasi-stellar objects (QSOs) as sources \citep[e.g.][]{Iliev:2005,Mesinger:2013}.
We vary the physical properties of the sources by changing the parameter $f_\gamma \equiv f_{\rm esc}\, f_\star \, N_i$ over simulation,
where $f_{\rm esc}$ is the escape fraction of ionizing photons,
$f_\star$ the star formation efficiency,
and $N_i$ the number of ionizing photons emitted per stellar baryon during its lifetime $\Delta t$ \citep[e.g.][]{Iliev:2006sw}.
$f_\gamma$ determines the source property in such a way that $f_\gamma
M_{\rm tot}$ photons are emitted from each grid cell during $\Delta t$,
where $M_{\rm tot}$ is the mass of halos inside the cell.
$\Delta t$ is also used as the time-step for finite-differencing
radiative transfer and rate equations. If
$\Delta t$ varies among simulations, a fixed $f_\gamma$ will generate
different numbers of photons accumulated. Therefore, it is sometimes
preferable to use a quantity somewhat blind to $\Delta t$; $g_\gamma
\equiv {f_\gamma}/(\Delta t / 10\,{\rm Myr})$ as defined in
\citet{Friedrich:2010} is such a quantity, which is basically the
emissivity of halos.

We also adopt two different types of halos:
low-mass ($10^8 \lesssim M / M_\odot \lesssim 10^9$) and high-mass ($M / M_\odot \gtrsim 10^9$) halos.
Both types are ``atomically-cooling'' halos,
where collisional excitation of the atomic hydrogen \Lya line predominantly initiates the cooling of baryons,
reaching the atomic-cooling temperature limit of $\sim 8000 \Kelvin$,
and is followed subsequently by H$_2$-cooling into a protostellar cloud.
One main difference exists in their vulnerability to external radiation:
in low-mass halos, when exposed to ionizing radiation,
the Jeans mass after photoheating overshoots their virial mass such that the gas collapse is prohibited,
while high-mass halos have a mass large enough to be unaffected by such photoheating.
We thus adopt a simple ``self-regulation'' criterion:
when a grid cell obtains $x > 0.1$, we fully suppress star formation inside low-mass halos in the cell \citep[e.g.][]{Iliev:2006sw}.
We also use in general different $f_\gamma$'s for different types,
when we include low-mass halos.
Cases 1 and 2 did not allow us to implement low-mass halos due to the limited mass resolution,
while Cases 3 and 4 did, thanks to the high mass resolution.

Finally, we note that halos having an even smaller mass range, $10^5 \lesssim M / M_\odot \lesssim 10^8$, or minihalos,
are not considered in this paper.
This can be justified because their impact on cosmic reionization,
especially at late stage, is believed to be negligible \citep[e.g.][]{Haiman:1999mn,Haiman:2006si}.
Nonetheless, the earliest stage of cosmic reionization must have started with minihalo sources,
most probably regulated by an inhomogeneous H$_2$ Lyman-Werner (LW) radiation background
(\citealt{Ahn:2008sq}; see also \citealt{Dijkstra:2008jk} for the impact of LW background on the growth of supermassive black holes).
A self-consistent simulation of cosmic reionization including minihalos and LW feedback has been achieved recently,
showing that about $\sim 20 \%$ of global ionization of the Universe can
be completed by minihalo sources at high-redshifts
(\citealt{Ahn:2012}).

\subsection{Calculation of the 21-cm Differential Brightness Temperature}\label{sec:sim_dtb}

The hyperfine splitting of the ground state of hydrogen leads to a transition with excitation temperature $T_* = 0.068 \Kelvin$.
The relative population of upper ($n_1$) and lower ($n_0$) states is determined by the spin temperature $T_S$, such that
\begin{equation}
\frac{n_1}{n_0} = 3 \fn{\exp}{-\frac{T_*}{T_S}}.
\end{equation}
Several pumping mechanisms determine $T_S$ as follows:
\begin{equation}
T_S = \frac{T_{\rm CMB} + y_\alpha T_\alpha + y_c T_K}{1 + y_\alpha + y_c} \,,
\end{equation}
where $T_{\rm CMB}$, $T_\alpha$, $T_K$, $y_\alpha$, and $y_c$ are the CMB temperature,
\Lya color temperature, the kinetic temperature of the gas,
the \Lya coupling coefficient, and the collisional coupling coefficient, respectively \citep{Purcell:1956,Field:1959}.

21-cm radiation is observed in emission against CMB if $T_S > T_{\rm CMB}$,
or in absorption if $T_S < T_{\rm CMB}$.
This is quantified by the 21-cm differential brightness temperature \citep[e.g.][]{Morales:2003vn}
\begin{equation}\label{eq:dtb_old}
\dtb = \frac{T_S - T_{\rm CMB}}{1+z} \left( 1 - e^{-\tau} \right) ,
\end{equation}
if 21-cm lines from neutral hydrogen at redshift $z$ are transferred toward us.
The optical depth $\tau$ is given by (see e.g. \citealt{Iliev:2002gj} for details)
\begin{eqnarray}\label{eq:tau}
\fn{\tau}{z} &=& (2.8 \times 10^{-4}) \fiducial{T_S}{1000 \Kelvin}{-1} \fiducial{h}{0.719}{} \fiducial{1+z}{10}{3/2} \nonumber \\
& & \times \fiducial{\Omega_b}{0.044}{} \fiducial{\Omega_m}{0.258}{-1/2} (1+\delta) .
\end{eqnarray}
$\dtb$ becomes appreciable enough to be detected by sensitive radio
antennae only when $T_S \gg T_{\rm CMB}$ or $T_S \ll T_{\rm CMB}$.
The former limit is believed to be reached quite early in the history of cosmic reionization,
because X-ray sources can easily heat up the gas over cosmological
scales with a small optical depth, such that $T_S \approx T_K \gg T_{\rm CMB}$ \citep[e.g.][]{Ciardi:2003hg}.
There exists another limiting case where $T_S$ is determined only by collisional pumping ($y_\alpha \ll 1$),
which is usually the case in the Cosmic Dark Ages as studied by
\citet{Shapiro:2006} and \citet{Kim:2010}, for instance. The latter limit
is reached when a strong Ly$\alpha$ background is built up while the
IGM is still colder than the CMB temperature, such that $T_S \approx T_K \gg T_{\rm CMB}$.
In this paper, we simply assume early heating of the IGM such that $T_S \gg
T_{\rm CMB}$. Equations~(\ref{eq:dtb_old}) and (\ref{eq:tau}) then give the limiting form of $\dtb$,
\begin{eqnarray}\label{eq:dtb}
\dtb &=& (28 \mK) \fiducial{h}{0.719}{} \fiducial{1+z}{10}{1/2} \nonumber \\
& & \times \fiducial{\Omega_b}{0.044}{}
\fiducial{\Omega_m}{0.258}{-1/2} (1+\delta) (1-x) .
\end{eqnarray}

We simply generate 3D maps of $\dtb$ on the radiative transfer grid over all redshifts using Equation~(\ref{eq:dtb}).
Note that in the limit $T_S \gg T_{\rm CMB}$, $\dtb \propto (1+\delta)(1-x)$ at a given redshift.
This limit, because $\dtb$ then becomes proportional to the underlying matter density,
allows one to perform cosmology without any bias,
by generating 3D maps of the matter density distribution as long as $x \ll 1$.
Of course, patchy evolution of H\,{\sc ii} bubbles will be strongly imprinted on this $\dtb$ map as well,
when the reionization process becomes mature \citep[e.g.][]{Shapiro:2013}.

\section{2D Genus of the 21-cm Differential Brightness Temperature}\label{sec:genus}

In 3D, the genus of a single connected surface is identical to the number of handles it has.
In 2D,
if a field of a variable $T$ is given,
the 2D genus for a threshold value, $T_{\rm th}$, is given by
\begin{equation}
g_{\rm 2D}(T_{\rm th}) \equiv \fn{N_+}{T_{\rm th}} - \fn{N_-}{T_{\rm th}},
\end{equation}
where $\fn{N_+}{T_{\rm th}}$ and $\fn{N_-}{T_{\rm th}}$ are the
number of connected regions with  $T > T_{\rm th}$ (hot spots) and
$T < T_{\rm th}$ (cold spots), respectively.
In general, as one varies $T_{\rm th}$,
$g_{\rm 2D}$ changes.
Therefore, $g_{\rm 2D}$ is a function of $T_{\rm th}$,
and its functional form is called the 2D genus. In our case, we
calculate the 2D genus curve of the field of the 21-cm differential brightness
temperature $\dtb$, by varying the threshold value $\delta T_{\rm b,\,th}$.

The 2D genus curve of a field can work as a characteristic of the field.
A useful template for this is a Gaussian random field,
because its 2D genus is given analytically by the relation
\begin{equation}
g_{\rm 2D,Gauss} \propto \nu \fn{\exp}{-\nu^2 / 2},
\label{eq:fg}
\end{equation}
where
\begin{equation}
\nu \equiv \frac{T_{\rm th} - \langle T \rangle }{\sigma_T}
\end{equation}
is the deviation from the average $\langle T \rangle$, in units of the standard deviation $\sigma_T$.
One can then compare the 2D genus curve of a given field to $g_{\rm 2D,Gauss}$
to see how close the field is to a Gaussian random field, for example.

The 2D genus is also equal to the integral of the curvature along the
contours of $T=T_{\rm th}$, divided by $2\pi$. This is because when a curvature is
integrated along a closed contour around a hot spot (cold spot),
its value is $2 \pi$ ($-2 \pi$). When a 2D field is pixelized, a
contour of $T=T_{\rm th}$ is composed of a series of line segments
with turns occurring at vertices.
On our rectangular 2D grid, every vertex shares four
pixels (except for vertices on
the edge and corner). Under these conditions, the
CONTOUR2D program is able to calculate the 2D genus by counting the
turning of a contour observed at each vertex of four pixels in an image
\citep{Melott:1989wv}: $1/4$ is contributing to the total genus from each vertex
with 1 hot pixel and 3 cold pixels, $-1/4$ from each vertex
with 3 hot pixels and 1 cold pixel, and zero otherwise.
 We do not calculate the modified 2D genus, $g_{\rm 2D,eff} \equiv
 g_{\rm 2D} - 2f$,
where $f$ is the areal fraction of hot spots on the sky.
$g_{\rm 2D,eff}$ is appropriate when applied to a full-sky field \citep{Gott:1989yj,Colley:2003sp,Gott:2006za},
while $g_{\rm 2D}$ is more appropriate for a relatively small,
restricted region on the sky.
Our simulation box is large enough to capture the typical size of H\,{\sc ii} bubbles ($\sim 10 - 20 \Mpc$ comoving)
during most phases of the cosmic reionization,
but still too small to cover the whole sky and capture fluctuations at much larger scales.

Our aim is to obtain the 2D genus curve from the frequency- and angle-averaged differential brightness temperature at given frequencies.
In order to compare different reionization models, we will choose
those frequencies such that different reionization simulations yield
the same global ionized fraction.
We will then vary the threshold $\delta T_{\rm b,\,th}$ to construct the 2D genus curve $\fn{g_{\rm 2D}}{\delta T_{\rm b,\,th}}$ for each case.

The amplitude of the 2D genus is roughly proportional to the field of
view at a fixed redshift. Throughout this paper, the 2D genus we
present is normalized to
the field of view of size (1 degree)$^2$.

\section{Results}\label{sec:res}

\subsection{Mock 21-cm Sky Maps}\label{sec:res_map}

\begin{figure*}[!t]
\centering
\includegraphics[width=0.9\textwidth]{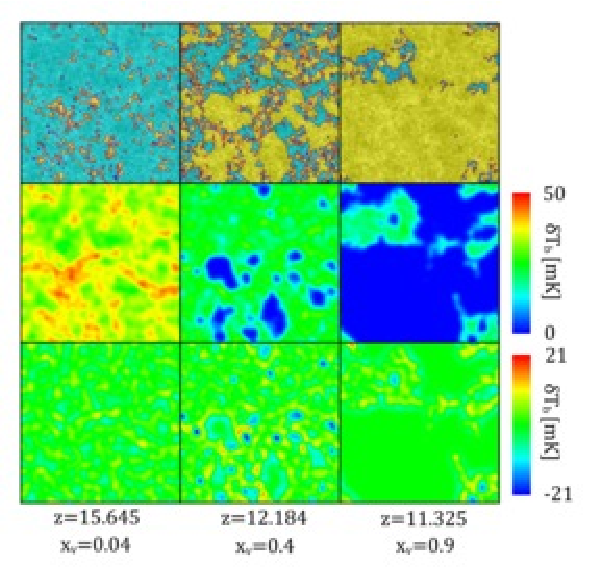}
\caption{21-cm maps at redshifts $z = 15.645, 12.184$ and 11.325 for f125\_1000S.
Top row: slices which are frequency-averaged into one band of $\Delta
\nu =0.2\MHz$, showing
ionized regions (yellow) and neutral regions (blue), superimposed on
the density field (light and dark for high-density and low-density
regions, respectively).
Middle row: corresponding 2D 21-cm differential brightness
temperature, averaged by a Gaussian beam of $\Delta \theta = 1'$.
Bottom row: same as middle, but with a compensated Gaussian beam of the same $\Delta \theta$.
\label{fig:map}}
\end{figure*}

Radio signals, observed by radio antennae, should be integrated over the angle, frequency and time
in order to achieve the sensitivity required for specific scientific goals.
Observing signals from high-redshifts is challenging, especially
because these signals are very weak while the required angle- and frequency-resolutions are relatively high.
Therefore, very intensive integration in time ($\sim 1000$ hours) is required for EOR observations in general.

We therefore choose the beam size and the frequency bandwidth
which we find adequate for generating distinctive 2D genus curves from different reionization scenarios,
under the assumption of very long time integration ($\sim 1000$ hours or more).
We generated mock 21-cm emission sky maps for our four reionization scenarios, f5, f250, f125\_125S and f125\_1000S,
averaging $\dtb$ on our computational grid over angle and frequency.
First, note that our box size corresponds to the frequency bandwidth
\begin{eqnarray}
\Delta \nu_{\rm box} &=&  \frac{\nu_0 L_{\rm box} H_0 \sqrt{\Omega_m (1+z)^3 + \Omega_\Lambda} }{c \, (1+z)^{2}} \nonumber \\
&\simeq & (4.10 \MHz) \fiducial{L_{\rm box}}{66 \Mpch}{} \fiducial{h}{0.719}{} \nonumber \\
& & \times \fiducial{\Omega_m}{0.258}{1/2} \fiducial{1+z}{15}{-1/2} \, ,
\end{eqnarray}
and the transverse angle
\begin{eqnarray}
\Delta \theta_{\rm box} &=& \frac{L_{\rm box}}{(1+z) \fn{D_A}{z}} = \frac{L_{\rm box}}{\fn{r}{z}} \nonumber \\
&=& (30.25 \arcmin) \fiducial{L_{\rm box}}{66 \Mpch}{} \nonumber \\
& & \times \fiducial{10563.5 - 11863.5 (1+z)^{-1/2}}{10563.5 - 11863.5
  (15)^{-1/2}}{-1} 
\label{eq:theta_box}
\end{eqnarray}
where $\nu_0 \equiv 1420 \MHz$ is the rest frame frequency of the line,
$\fn{D_A}{z}$ the angular diameter distance,
$L_{\rm box}$ the comoving length of our simulation box,
$c$ the speed of light
and $H_0 \equiv 100\,h \kmsMpc$ the Hubble parameter at
$z=0$. The numerator in the last parenthesis in Equation~(\ref{eq:theta_box}) is
a fitting formula for the line-of-sight comoving distance $r(z)$
in units of Mpc, under the cosmological parameters used in this paper.

We first calculate the bare differential brightness temperature on our computational grid by using Equation~(\ref{eq:dtb}).
We then integrated them over frequency with chosen $\Delta \nu$'s.
We also consider the Doppler shift due to peculiar velocity.
Along the line-of-sight (LOS), the frequency-averaged $\dtb$ then becomes
\begin{equation}
\langle \dtb \rangle _{\Delta \nu} = \frac{\sum_i (\dtb)_i f_i}{N_{\Delta \nu}} \, ,
\end{equation}
where $N_{\Delta \nu} \equiv \fn{\Delta L}{\Delta \nu} / l_{\rm cell}$ is the number of cells corresponding to $\Delta \nu$ at a given redshift,
$f_i$ the fraction of a cell $i$ entering the band after the Doppler shift,
\begin{eqnarray}
\fn{\Delta L}{\Delta \nu} & \equiv & \frac{\Delta \nu \, c \, (1+z)^{2}}{\nu_0 H_0 \sqrt{\Omega_m (1+z)^3 + \Omega_\Lambda}} \nonumber \\
&\simeq & (3.22 \Mpch) \fiducial{\Delta \nu}{0.2 \MHz}{} \fiducial{h}{0.719}{-1} \nonumber \\
& & \times \fiducial{\Omega_m}{0.258}{-1/2} \fiducial{1+z}{15}{1/2} \,
\end{eqnarray}
the comoving length along LOS corresponding to $\Delta \nu$,
and $l_{\rm cell}$ the comoving length of a unit cell.
For statistical reasons, we generate four 21-cm emission maps for each LOS direction from a single simulation box
by periodically shifting the box by a quarter of box size
($L_{\rm box} / 4$). These maps will be almost mutually
independent, because $L_{\rm box} / 4 \gg \fn{\Delta L}{\Delta \nu}$.
We also take three different directions for LOS ($x$-, $y$-, and $z$-direction) by rotating the box,
such that we have in total 12 (almost) independent 21-cm sky maps from a single simulation box at a given redshift.

Because the shape of the beam varies over different antennae,
we choose two different types of beams to average the signal in angle: Gaussian and compensated Gaussian.
We first convolve the 21-cm emission map from simulations with a Gaussian filter
\begin{equation}
\fn{W_{\rm G}}{\theta} = \frac{1}{2 \pi \sigma^2} \, \fn{\exp}{-\frac{\theta^2}{2 \sigma^2} }
\end{equation}
with a full width at half maximum (FWHM) $\Delta \theta = 2 \sigma \sqrt{ 2 \ln 2}$.
We also use a compensated Gaussian filter given by
\begin{equation}
\fn{W_{\rm CG}}{\theta} = \frac{1}{2 \pi \sigma^2} \left( 1 - \frac{\theta^2}{2\sigma^2} \right) \fn{\exp}{1 - \frac{\theta^2}{2\sigma^2} }
\end{equation}
with a FWHM given by
\begin{equation}
\Delta \theta = 2 \sigma \sqrt{2 (1 - \fn{\mathrm{LambertW}}{e/2} )}
\end{equation}
where $\fn{\mathrm{LambertW}}{e/2} \approx 0.685$.

A Gaussian filter is widely used in the literature to approximate actual
beams. In general, an interferometer will
not retain the average (large scale) signal, but only record the
fluctuations between minimum and maximum angular scales (depending on
the largest and smallest baselines). As the largest scales to which
the upcoming interferometers are sensitive are larger than our
simulated images, a reasonable approach would be to subtract the mean
signal from our images. However, for the determination of the genus
this does not make a difference other than shifting the threshold value, so we retain the average signal here.
A compensated Gaussian filter roughly mimics the beam of a compact interferometer \citep[e.g.][]{Mellema:2006pd},
even though real beams have sometimes much more complicated shapes
depending on the actual configuration of the antenna.
We just take $W_{\rm CG}$ as an extreme case for ``dirty'' beams,
clearly distinct from $W_{\rm G}$, because this filter is somewhat
pathological as it suppresses both small and large scale features.
When $\theta > \sqrt{2} \sigma$,
$\fn{W_{\rm CG}}{\theta} < 0$ such that a data field convolved with $\fn{W_{\rm CG}}{\theta}$ may change its sign from point to point.
21-cm signals filtered this way would show a seemingly unnatural H\,{\sc ii} bubble feature,
or in the worse case, even absorption signals ($\dtb < 0$),
although Equation~(\ref{eq:dtb}) implies that $\dtb > 0$ when $T_S
\gg T_{\rm CMB}$.

Figure~\ref{fig:map} shows how the actual 2D $\dtb$ field (top rows) will be observed in different filtering schemes
(middle: $W_{\rm G}$; bottom: $W_{\rm CG}$).
Three different epochs were chosen to represent early (volume weighted
ionized fraction $x_{\rm v} = 0.04$),
middle ($x_{\rm v} = 0.4$), and late ($x_{\rm v} = 0.9$) stages.
$W_{\rm C}$ generates reasonably filtered $\dtb$ maps at all the three epochs.
At very early ($x_{\rm v} = 0.04$) and late ($x_{\rm v} = 0.9$) stages,
fluctuations in $\dtb$ are dominated by $1+\delta$ and $1-x$, respectively,
while at the middle ($x_{\rm v} = 0.4$), both $1+\delta$ and $1-x$ contribute to the fluctuations.

The compensated Gaussian beam adds ripple structures to the emission signals.
When a uniform field confined within a certain boundary is convolved with $W_{\rm CG}$,
this ripple becomes visible only along the boundary.
As the ionized regions grow, therefore,
a compensated Gaussian beam produces a negative through of width similar to the beam FWHM at the boundaries of the bubbles.
In general, $W_{\rm CG}$ changes the topology of the 21-cm signal.
For most of our analysis, therefore, we just use $W_{\rm G}$.
Further comparison between these two filters will be made
in Section~\ref{sec:res_genus_evol}.

\subsection{Evolution, Sensitivity, Bubble Distribution, and Power Spectrum}\label{sec:res_general}

\begin{figure*}[!t]
\centering
\includegraphics[width=0.32\textwidth]{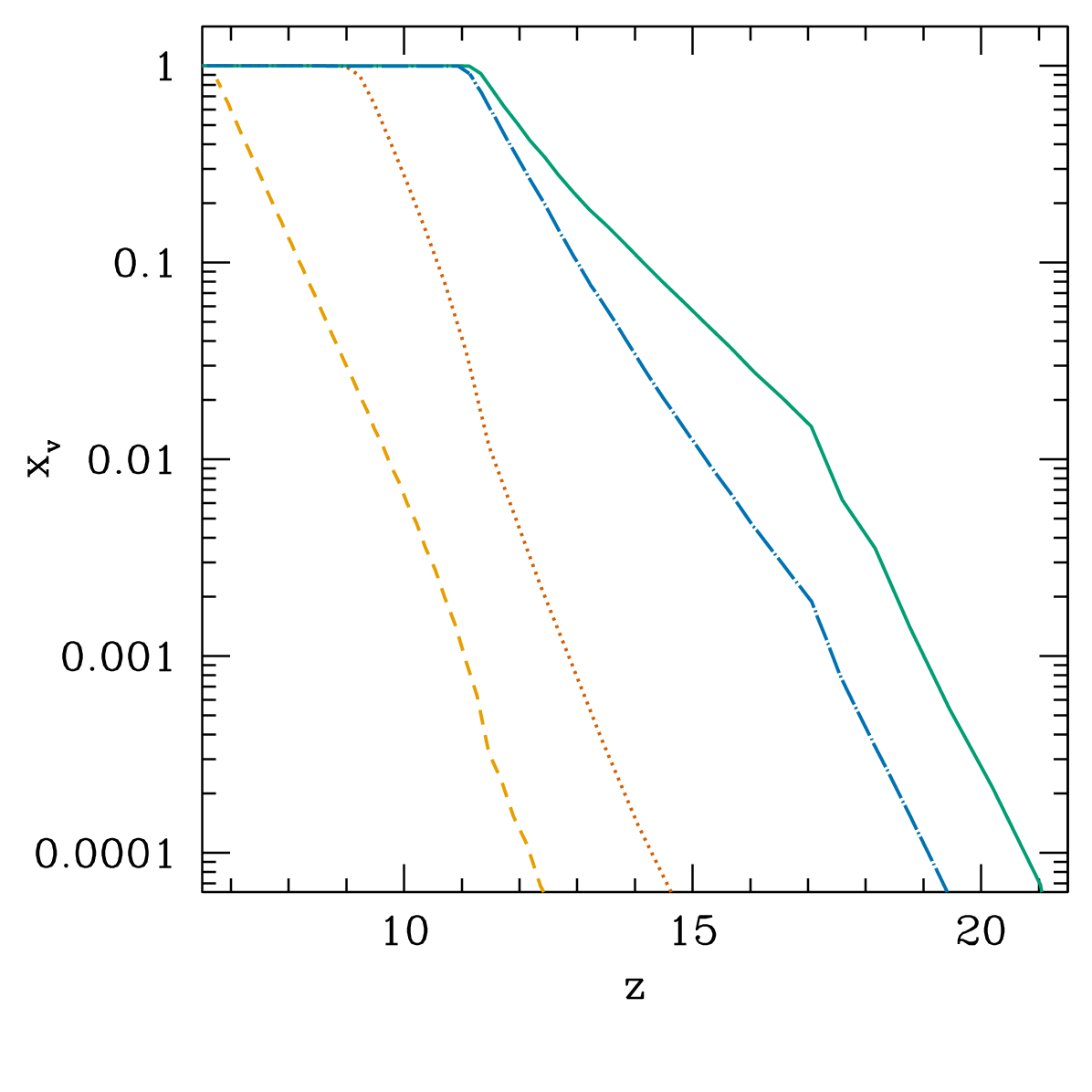}
\includegraphics[width=0.32\textwidth]{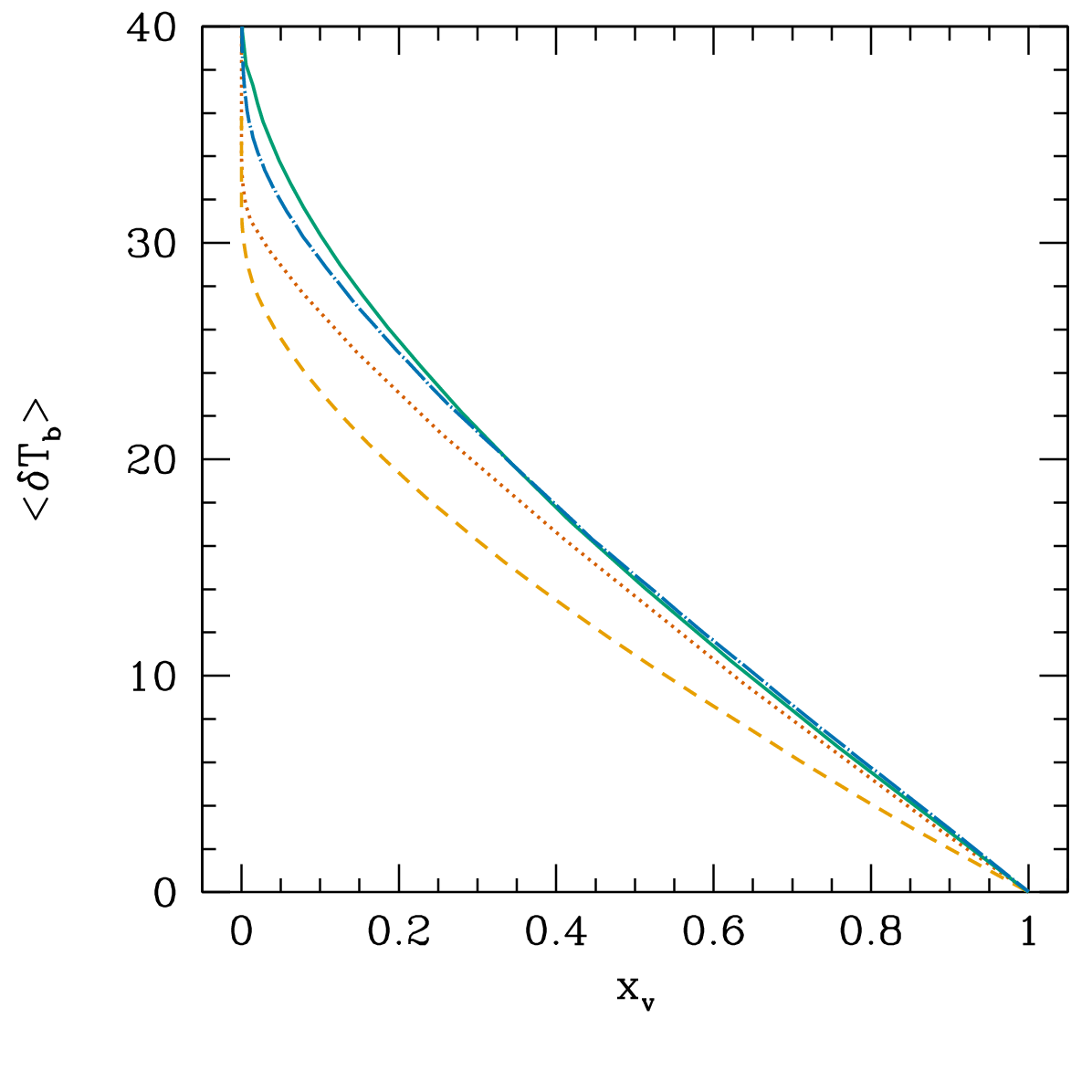}
\includegraphics[width=0.32\textwidth]{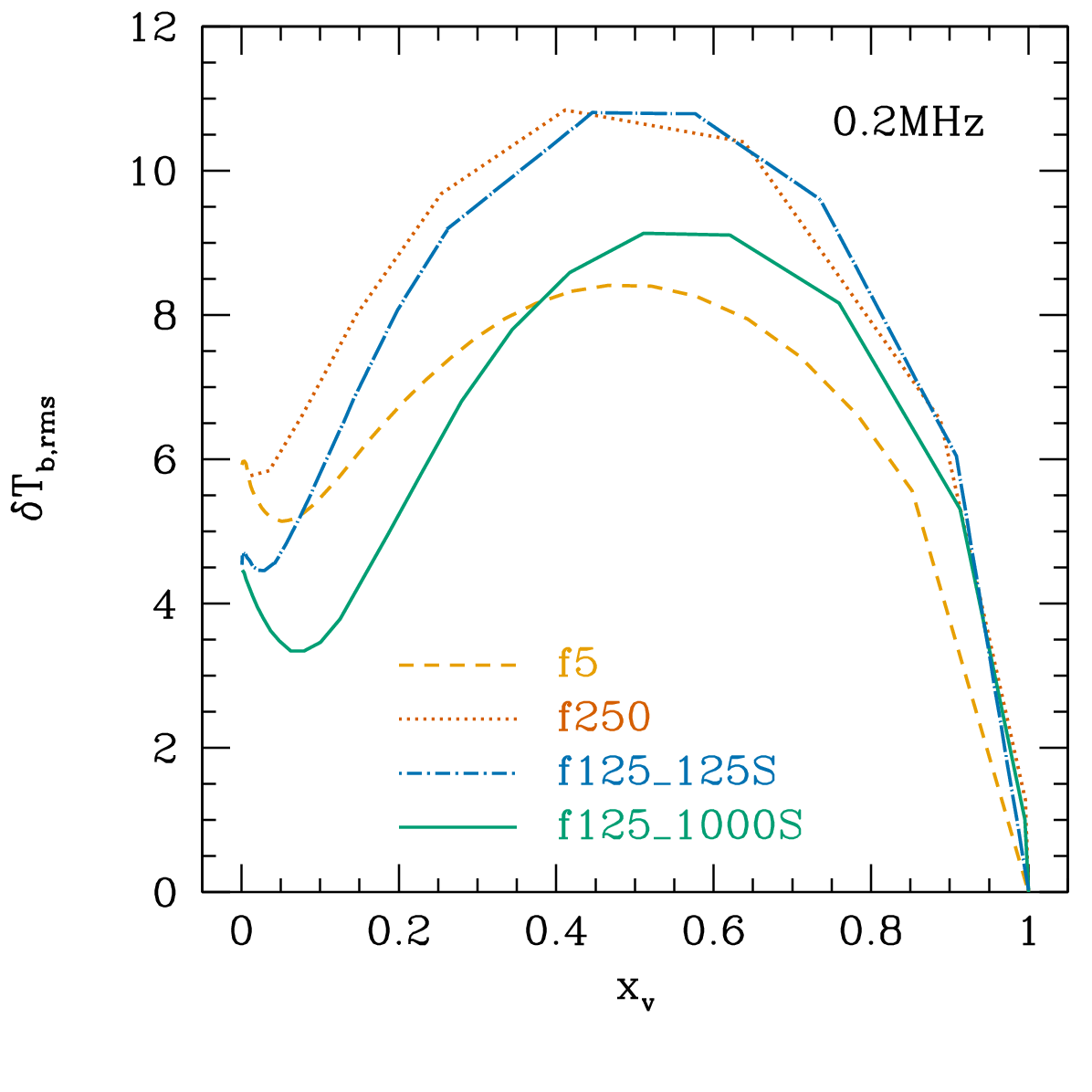}
\caption{Left: Evolution of the mean volume ionized fraction as a function of redshift.
Middle: mean differential brightness temperature with $1\arcmin$ FWHM Gaussian beam and $0.2\MHz$ bandwidth, as a function of the mean volume ionized fraction.
Right: rms fluctuation of the differential brightness temperatures
filtered by the same beam and bandwidth, as a function of the mean volume ionized fraction.\label{fig:ion_mean_rms}}
\end{figure*}

The evolution of the global volume-weighted ionized fraction
$x_{\rm v}$ varies significantly over different reionization scenarios,
as seen in Figure~\ref{fig:ion_mean_rms}.
First, the end of reionization in the f5 and f250 cases occurs relatively later than in the f125\_125S and f125\_1000S cases,
because these are tied only to high-mass ($M > 2 \times 10^9 M_\odot$) halos,
which collapse much later than low-mass ($10^8 \lesssim M / M_\odot \lesssim 10^9$) halos.
Second, f5 and f250 also show a steeper evolution in $x_{\rm v}$,
because there is no chance for slow evolution of the total luminosity from self-regulation of small-mass halos.

In general, it is a better practice to compare different models at a fixed global ionized fraction rather than at a fixed redshift,
partly because there is still too much freedom in the exact epoch of the end of reionization,
the time-integrated optical depth to Thompson scattering of CMB
photons, etc. More importantly, a fixed global ionized fraction among
different models is reached by roughly the same number of H-ionizing
photons accumulated from the beginning of the EOR.
We therefore use $x_{\rm v}$ as the time indicator of the global evolution throughout this paper.

Using $x_{\rm v}$ as the time indicator, we first observe similar trends
in the evolution of $\dtb$ among different models.
The mean differential brightness temperature $\langle\dtb\rangle$ starts from $\mdtb \approx 30 - 40 \mK$ and gradually decreases in time,
as seen in Figure~\ref{fig:ion_mean_rms}.
The root-mean-square (rms) fluctuation of $\dtb$, $\rdtb$, reaches the maximum ($\sim 8 - 11\mK$) at $x_{\rm v} \approx 0.4 - 0.6$ (Figure~\ref{fig:ion_mean_rms}),
when the signal is filtered with $\Delta \nu = 0.2 \MHz$ and $\Delta \theta = 1 \arcmin$.

\begin{figure*}[!t]
\centering
\includegraphics[width=0.45\textwidth]{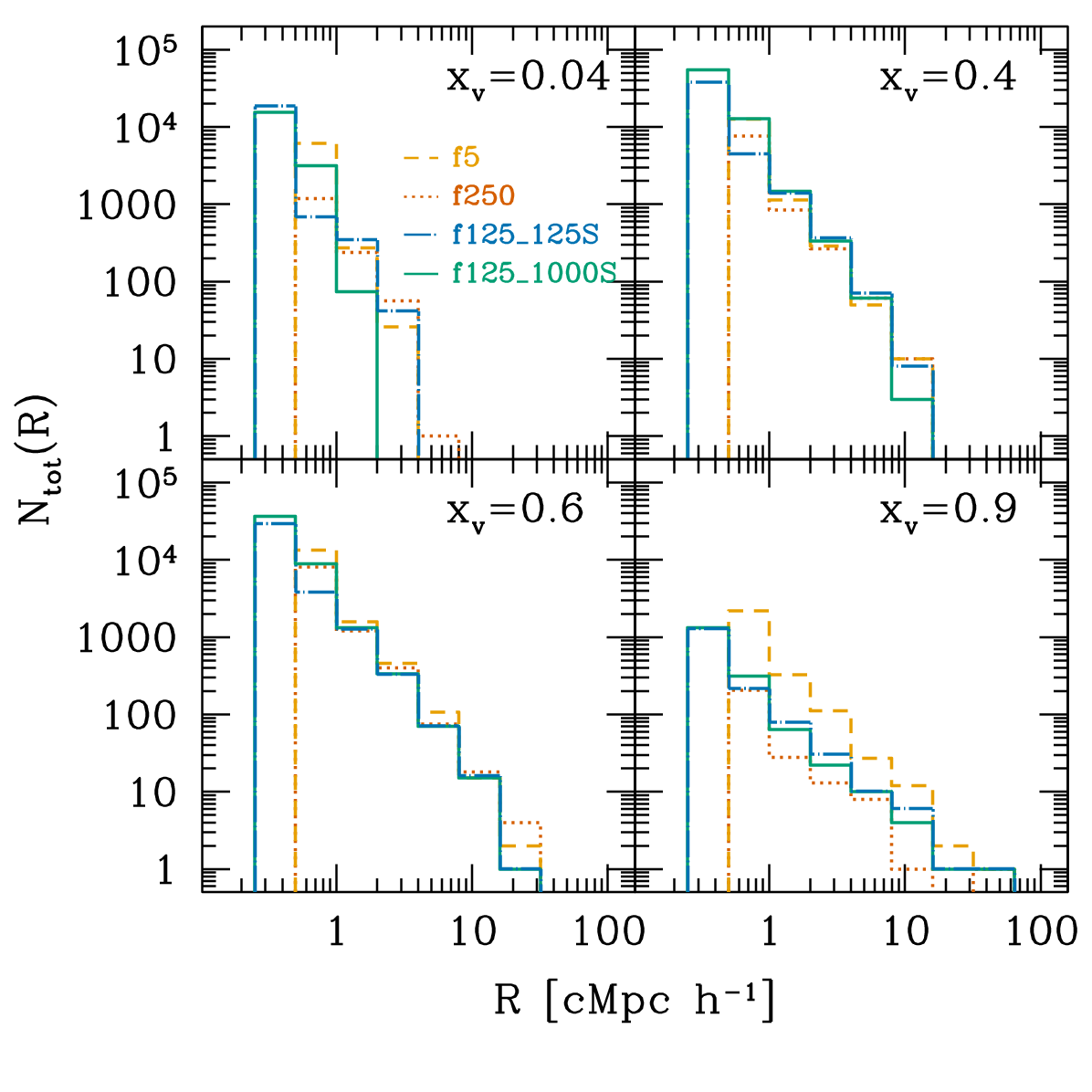}
\includegraphics[width=0.45\textwidth]{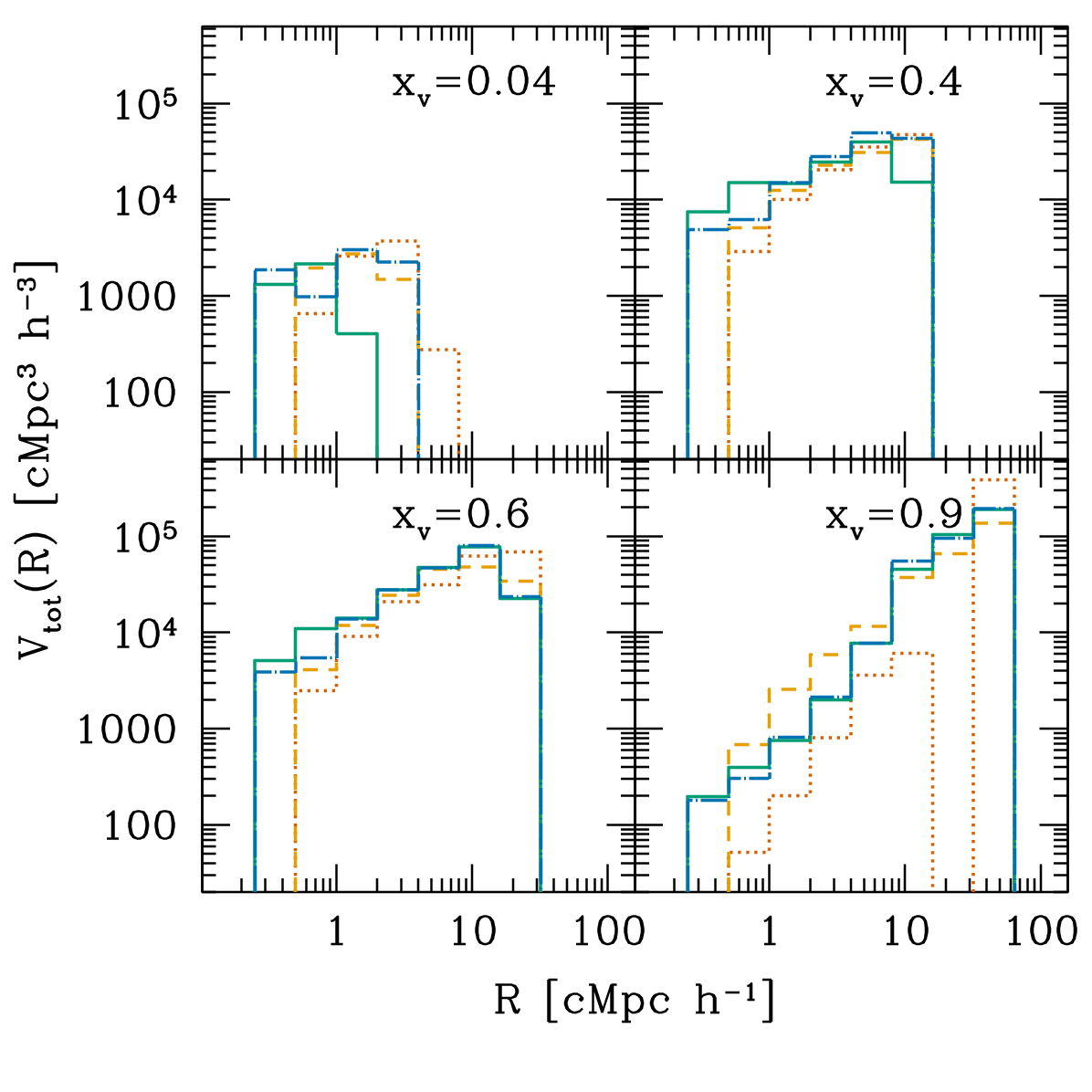}
\caption{
Left: Probability distribution functions of the bubble size.
We determined the size of a bubble using the method of \citet{Zahn:2006sg},
and select mutually exclusive bubbles using that of \citet{Hoyle:2001kn}.
Right: volume-weighted probability distribution functions of the bubble size.\label{fig:mcquinn}}
\end{figure*}

Albeit similar trends, a detailed analysis will be needed to
discriminate between different models and ultimately probe the
properties of the sources.
 The probability distribution function (pdf) of the bubble size is a
 useful tool for understanding the impact of these source properties.
The bubble size is associated with the mass spectrum of halos and the
ionizing efficiency $f_\gamma$, simply because it roughly reflects the total
number of H-ionizing photons emitted into the bubble.
In order to find the pdf of the bubble size, we use a hybrid method.
First, the size of a bubble is determined by the method in \citet{Zahn:2006sg}:
the bubble size is the maximum radius of a sphere from each simulation cell inside which the ionized fraction is over 90\%.
In this way, every cell in the box is associated with a bubble with a certain size.
We then use the void-finding method of \citet{Hoyle:2001kn},
which has been extensively used for finding cosmological voids which are mutually exclusive in space.
The bubbles are sorted in size from the largest to the smallest.
The largest bubble is considered an isolated one.
We then move to the next largest bubble and check if there is any overlap in volume with the fist one.
We iterate this over the bubble list until the overlap becomes less than 10\%,
which then registers another isolated bubble.
This process is further iterated and we obtain the full list of mutually exclusive bubbles (to the extent of a 10\% overlap).

In Figure~\ref{fig:mcquinn}, we show the pdf of the bubble size obtained in
this way.
We can observe a clear distinction of f5 and f250 from f125\_125S and f125\_1000S.
In f5 and f250, a given ionized fraction is reached only by high-mass halos.
For example, to reach $x_{\rm v} \sim 0.6$,
the evolutionary stage of cosmological structure formation should enter a highly nonlinear
phase because there is only one species (high-mass halos) responsible
for such high $x_{\rm v}$.
In contrast, in f125\_125S and f125\_1000S,
both the two different species (high- and low-mass halos) contribute to, for example, $x_{\rm v} \sim 0.6$.
Therefore, structure formation will be in a less nonlinear stage than the former cases.
Correspondingly, we can expect stronger clustering of sources for f5 and f250,
and weaker for f125\_125S and f125\_1000S.
Correspondingly, the merger of bubbles would be stronger for the former and weaker for the latter, respectively.
This is finally reflected in the relative contribution to a given $x_{\rm v}$ from largest bubbles,
as seen in Figure~\ref{fig:mcquinn}.
Large bubbles dominate $x_{\rm v}$ in f5 and f250,
while there is a relatively smaller contribution to $x_{\rm v}$ in f125\_125S and f125\_1000S.

\begin{figure}[!t]
\includegraphics[width=0.45\textwidth]{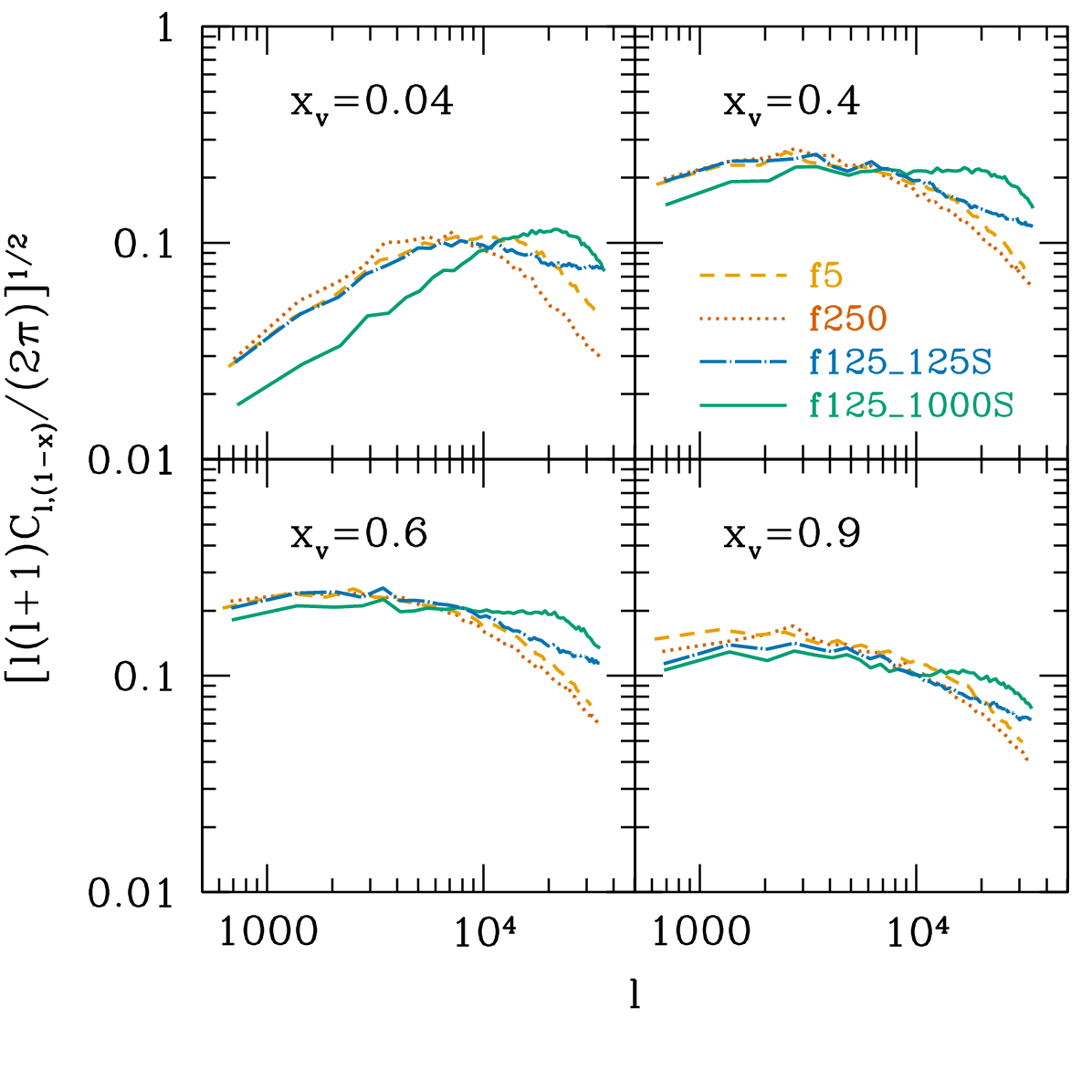}
\caption{
2D angular power spectra of the neutral fraction $1-x$.\label{fig:power}}
\end{figure}
We also obtain 2D angular power spectra of the neutral fraction, $1-x$, for further analysis.
This provides insights on the clustering scale of the sources of radiation, or H\,{sc ii} bubbles,
through the location of peaks in the power spectrum.
We follow the convention used in CMB analysis and constructed $[\ell (\ell +1) C_\ell / 2 \pi]^{1/2}$ in spherical harmonics,
where $\ell$ roughly corresponds to $2\pi / \theta$
(see Figure~\ref{fig:power}).
At $x_{\rm v} = 0.04$, it is clearly seen how cosmic reionization commences in each scenario.
f5 and f250 show peaks in $\ell (\ell +1) C_\ell$ at $\ell \sim 10^4$,
or at the comoving length scale $x \sim 6 - 7 \Mpch$,
while f125\_1000S shows a peak at $\ell \sim 3 \times 10^4$, or at $x \sim 2 \Mpch$,
and f125\_125S shows a long uniform tail from $\ell \sim 3 \times 10^4$ to $\ell \sim 4 \times 10^4$.
The behavior of 2D angular power spectra indicates stronger merger of (otherwise) individual bubbles into a larger scale at $x \sim 6 - 7 \Mpch$ in f5, f250 and f125\_125S than in f125\_1000S.
Even though f125\_125S has a tail up to $\ell \sim 4 \times 10^4$ ($x \sim 1 \Mpch$),
a smaller $f_\gamma (=125)$ in small halos than that of f125\_1000S ($f_\gamma = 1000$) makes these small halos less efficient.
This is true at all redshifts:
the power spectrum of f125\_125S is hardly distinguishable from those of f5 and f250 at any time,
except for higher power in the smallest length scale (highest $\ell$).
In contrast, a distinctive clustering scale, $\ell \sim 3 \times 10^4$ ($x \sim 2 \Mpch$),
is shown at all redshifts in f125\_1000S, due to the clustering  of
small halos of the highest efficiency.

In short, both the bubble pdf and the 2D angular power spectrum of the neutral fraction may be useful tools
for understanding the nature of different reionization scenarios.
In Section~\ref{sec:res_genus},
when we analyze 2D genus curves of different cases,
we will characterize these properties to see whether 2D genus curves created from possible observations can show similar fundamental nature as well.

\subsection{Genus Properties}\label{sec:res_genus}

We generate $\fn{g_{\rm 2D}}{\dtb}$ from selected epochs at which $x_{\rm v} = 0.04$, 0.4, 0.6 and 0.9 as described in Section~\ref{sec:genus}.
The base 2D field of $\dtb$ is averaged in frequency with $\Delta \nu = 0.2$, 1 and $2 \MHz$,
and in angle with $\Delta \theta = 1\arcmin$, $2\arcmin$ and $3\arcmin$.
We vary $\Delta \nu$ and $\Delta \theta$ to quantify the competing effects of changing resolution and sensitivity,
and also to make comparison with characteristics of future radio antennae.
In angle-averaging, only $W_{\rm G}$ is used in all the cases,
except for f125\_1000S where $W_{\rm CG}$ is also applied to understand the impact of the beam shape on $g_{\rm 2D}$.

\subsubsection{2D Genus of Density Fluctuations}\label{sec:res_genus_matt}
\begin{figure}[!t]
\includegraphics[width=0.45\textwidth]{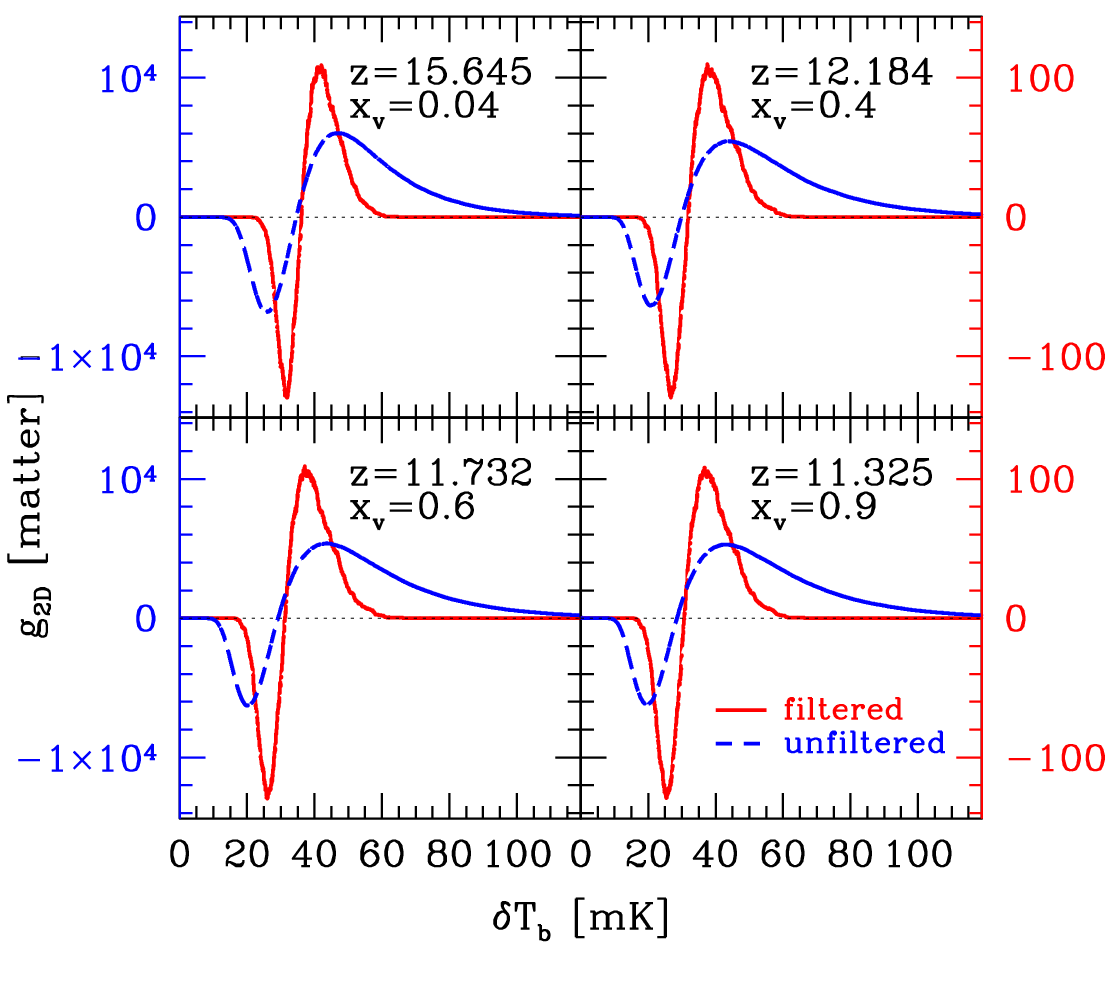}
\caption{2D genus defined by the threshold differential brightness temperature within a
  $\sim 30\arcmin \times 30\arcmin$ square (corresponding
  to our box size) in the sky  assuming all the gas was neutral,
filtered with $1\arcmin$ FWHM Gaussian beam and $0.2\MHz$ bandwidth (red).
In comparison, we also show the 2D genus of the differential
brightness temperature with the same condition but
from one-cell thick slice in our simulation
(blue). $\fn{g_{\rm 2D}}{\dtbn}$ of the bare $N$-body field is
thus subject to smoothing over the size of the cell.
This result is based upon the high-resolution N-body simulation (used
for Cases 3 and 4). \label{fig:genus_matter}}
\end{figure}

Is there any useful template genus curve to be compared with $\fn{g_{\rm 2D}}{\dtb}$?
There is such a template indeed,
namely $g_{\rm 2D}$ of an artificial field of $\dtb$, which reflects the density fluctuation only.
Let us denote this quantity by
\begin{eqnarray}
\dtbn &\equiv & \fn{\dtb}{x = 0} \nonumber \\
&=& (28 \mK) \fiducial{h}{0.719}{} \fiducial{1+z}{10}{1/2} \nonumber \\
& & \times \fiducial{\Omega_b}{0.044}{} \fiducial{\Omega_m}{0.258}{-1/2} (1+\delta).
\end{eqnarray}
The distribution of $\dtbn$ will be Gaussian in the linear regime ($|\delta| \ll 1$),
which will provide a well-defined, analytical $\fn{g_{\rm 2D, Gauss}}{\dtbn}$.
Even when $\fn{g_{\rm 2D}}{\dtbn}$ starts to deviate from $\fn{g_{\rm 2D, Gauss}}{\dtbn}$
due to nonlinear evolution of high density peaks,
$\fn{g_{\rm 2D}}{\dtbn}$ will be used to indicate overdense
($\delta > 0$) and underdense ($\delta < 0$) regions.
Figure~\ref{fig:genus_matter} implies that $\dtbn >0$ corresponds to
$\delta > 0$, while $\dtbn <0$ to $\delta <0$.

The curve $\fn{g_{\rm 2D}}{\dtbn}$, plotted in Figure~\ref{fig:genus_matter},
shows how the matter density evolves.
At $z \approx 10 - 15$, there already exist extended tails into high $\dtbn$,
because these correspond to high-density peaks that have evolved into the nonlinear regime.
Note that a purely Gaussian random field generates $g_{\rm 2D,Gauss}$,
which is symmetrical around $\nu = 0$,
while high-density peaks (on high $\nu$) that evolved nonlinearly deviate from the Gaussian distribution
to make $g_{\rm 2D}$ deviate from $g_{\rm 2D,Gauss}$.
Because we will use filtered $\dtb$ maps,
$g_{\rm 2D}$ of filtered $\dtbn$ will be the template to be used throughout this paper.
Filtering makes both the amplitude and width of $\fn{g_{\rm 2D}}{\dtbn}$ shrink (Figure~\ref{fig:genus_matter}).
Nevertheless, even after filtering,
regions with positive $g_{\rm 2D}$ correspond to overdense regions and negative $g_{\rm 2D}$ to underdense regions,
because the mean-density point (where the $\fn{g_{\rm 2D}}{\dtbn}$ curve crosses the $\dtb$ axis)
is almost unshifted.
This will work as a perfect indicator, if comparison is made with $\fn{g_{\rm 2D}}{\dtb}$,
about how different the $\dtb$ field is from the underlying density field.
$\fn{g_{\rm 2D}}{\dtbn}$ can be obtained accurately for any redshift as long as cosmological parameters are known to high accuracy,
which is the case in this era of precision cosmology.

\subsubsection{Evolution of $g_{\rm 2D}$ and Beam Shape Impact}\label{sec:res_genus_evol}
\begin{figure*}[!t]
\centering
\includegraphics[width=0.45\textwidth]{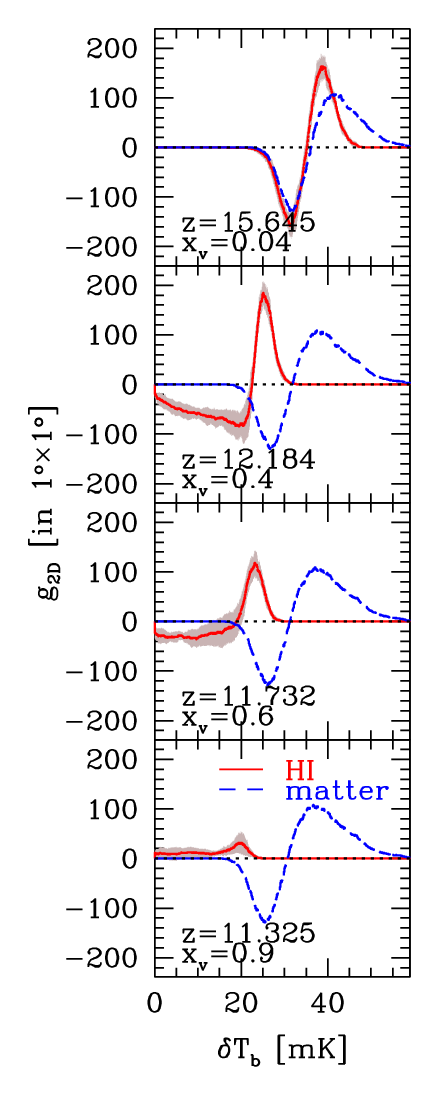}
\includegraphics[width=0.45\textwidth]{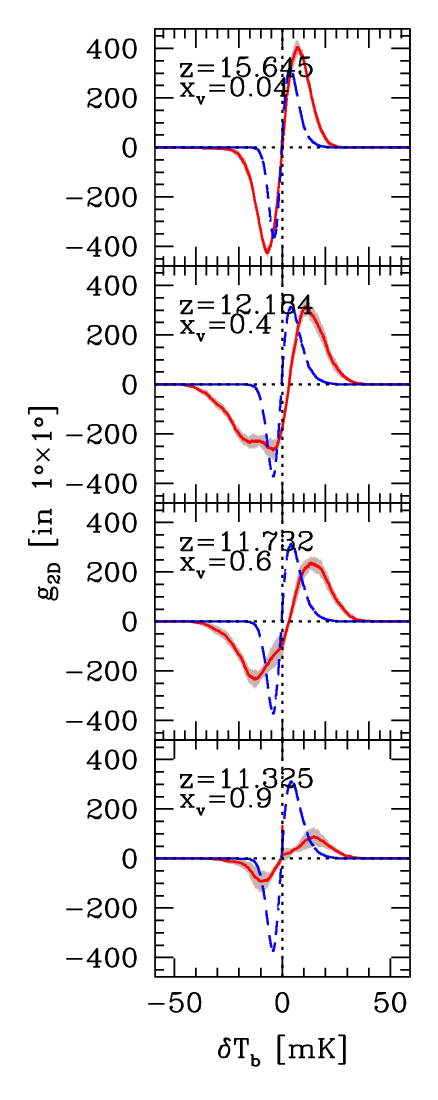}
\caption{Left: 2D genus of $\dtb$ for f125\_1000S, filtered
  with a $1\arcmin$ FWHM
  Gaussian beam and $0.2\MHz$ bandwidth (solid), compared to
  the template 2D genus out of the underlying density field (dashed),
  also filtered in the same way.
  Right: Same as the left panel, but with a compensated Gaussian beam applied.
  \label{fig:genus_matt_dTb}}
\end{figure*}

We found that the evolution of $\fn{g_{\rm 2D}}{\dtb}$ contained both generic and model-dependent features.
We describe the generic features based on the f125\_1000S case.
We will  describe the model-dependent features in Section~\ref{sec:res_genus_model}.

We find that the change of $\fn{g_{\rm 2D}}{\dtb}$ clearly shows how reionization proceeds in time.
The left panel of Figure~\ref{fig:genus_matt_dTb} shows
$W_{\rm G}$-filtered $\fn{g_{\rm 2D}}{\dtbn}$ of the underlying
matter density field (blue dashed line) and $\fn{g_{\rm 2D}}{\dtb}$
of the differential brightness temperature field of the HI gas (red line).
In the earliest phase of reionization ($x_{\rm v} = 0.04$),
$\fn{g_{\rm 2D}}{\dtb}$ is much smaller than
$\fn{g_{\rm 2D}}{\dtbn}$ at high temperature thresholds ($\dtbn \gtrsim 50 \mK$).
It is because the highest density peaks have been ionized by the
sources inside them, and are dropped out of the neutral gas
distribution.
It can also be noted that the amplitudes of both the maximum (at $\dtb
\sim 38 \mK$) and minimum (at $\dtb \sim 33 \mK$) of
$\fn{g_{\rm 2D}}{\dtb}$ are higher than those of $\fn{g_{\rm 2D}}{\dtbn}$.
The birth of new islands and peninsulas made of H I regions --- some peninsulas
may appear as
islands at certain $\dtbth$ --- is responsible for the former,
while the birth of new lakes made of H II regions is responsible for the latter.
The birth of islands and peninsulas occurs at mildly overdense regions ($\dtbn \sim 38 \mK$)
because bubbles are clustered such that some of them merge with one another to fully or partly surround neutral regions.
Note that a neutral region identified as an island in 2D may well be a
cross-section of a vast neutral region in 3D.

When the Universe enters the intermediate phase of reionization ($x_{\rm v} \approx 0.4 - 0.6$),
almost all overdense regions ($\dtbn \gtrsim 30 \mK$) have been ionized to form larger bubbles.
This is reflected in the relatively low amplitude trough of $\fn{g_{\rm 2D}}{\dtb}$,
which appears in very low-density ($\dtbn \lesssim 20 \mK$) regions.
Now many more islands appear in mildly underdense ($\dtbn \sim 25 \mK$) regions
because larger and more clustered bubbles penetrate further into the low-density IGM
and are more efficient in forming new islands.
As time passes from $x_{\rm v} = 0.4$ to $x_{\rm v} = 0.6$, the
amplitude of $g_{\rm 2D}$  decreases as
bubbles merge with each other and neutral clumps disappear.
Further penetration of bubbles into the lower-density IGM from $x_{\rm v} = 0.4$ to $x_{\rm v} = 0.6$ is
also reflected in the maximum value of $\dtb$ for nonzero $\fn{g_{\rm 2D}}{\dtb}$.

Finally, in the late stage ($x_{\rm v} = 0.9$),
all ionized bubbles have been connected and the last surviving neutral
clumps exist to give positive
$g_{\rm 2D}$ at $\dtb \sim 20 \mK$.
Note that these clumps exist only in underdense regions,
which is a clear indication of the fact that reionization proceeds in an inside-out fashion:
high-density regions will be ionized first due to the proximity to the sources of radiation,
and low-density regions will be ionized later.

We then briefly investigate the impact of beam shapes.
The impact of $W_{\rm CG}$ on $g_{\rm 2D}$ is demonstrated on the right panel of Figure~\ref{fig:genus_matt_dTb}.
Note that the Fourier transform of $W_{\rm CG}$ is proportional to
$k^2 \sigma^2 \exp(-k^2 \sigma^2 /2)$. From the convolution theorem,
it is obvious that the filtered field
becomes the $W_{\rm G}$-smoothed, negative Laplacian ($-\nabla^2$) of the
field. Note that Laplacian is equivalent to the divergence of the
gradient.
Therefore, if the $\dtb$ field is filtered by $W_{\rm CG}$,
$\dtb$s in those regions with the steepest spatial gradient in $x$ or
$\delta$ will correspond to extrema. We also note that the sign of the
Laplacian depends on the morphology of $x$: if H II bubbles form
in the sea of neutral gas, gradients of $\dtb$ diverge ($-\nabla^2
\dtb <0$), while if
neutral clumps remain in the sea of ionized gas, gradients of $\dtb$
converge ($-\nabla^2 \dtb >0$). An interesting feature in
$\fn{g_{\rm 2D}}{\dtb}$ is indeed observed to
move from regions with $\dtb <0$ at a relatively early ($x_v \sim 0.4$)
stage to regions with $\dtb >0$ at a later ($x_v \sim 0.9$)
stage. Nevertheless, because the topology of the 21-cm signal is
processed further (Gaussian smoothing and Laplacian) in the case of $W_{\rm CG}$ than that of
$W_{\rm G}$ (Gaussian smoothing only), the interpretation becomes
less transparent. We also note that $W_{\rm CG}$ filters out almost
all the scales except for the filtering scale, while $W_{\rm G}$
filters out only scales smaller than the filtering scale.

The nice interpretative power of $\fn{g_{\rm 2D}}{\dtb}$ is
somewhat lost with $W_{\rm CG}$.
Based on this, we suggest that any artificial effects from ``dirty'' beams should be removed or minimized in real observations.
Only when the effective beam shape becomes something close to $W_{\rm G}$,
general analyses including our genus method will exploit their full potential.

\subsubsection{Model-Dependent Feature and Required Sensitivity}\label{sec:res_genus_model}
We finally investigate whether our 2D genus analysis can discriminate between
different reionization scenarios. Ultimately, if this turns out to be true,
we may be able to probe properties of radiation sources at least indirectly, because these
properties determine how cosmic reionization proceeds.

\begin{figure*}[!t]
\centering
\includegraphics[width=0.9\textwidth]{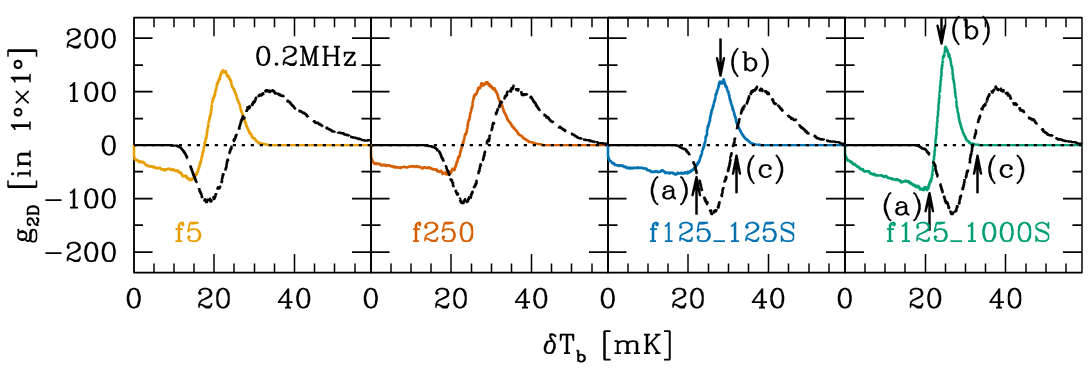}
\includegraphics[width=0.9\textwidth]{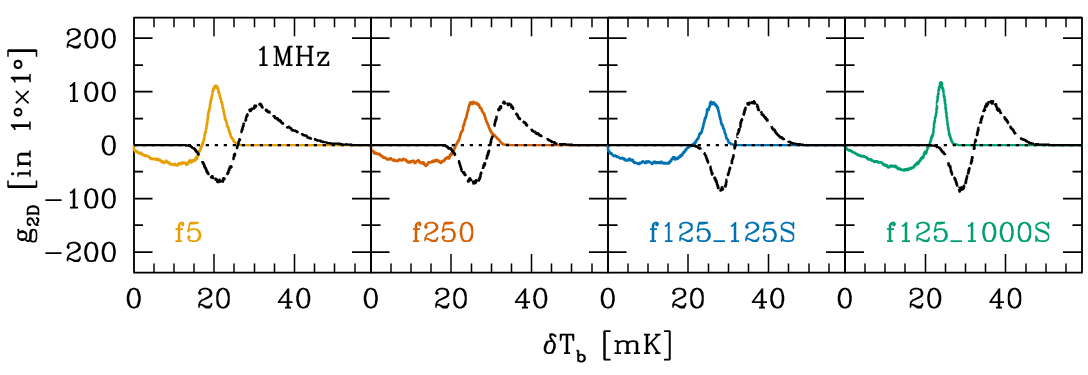}
\includegraphics[width=0.9\textwidth]{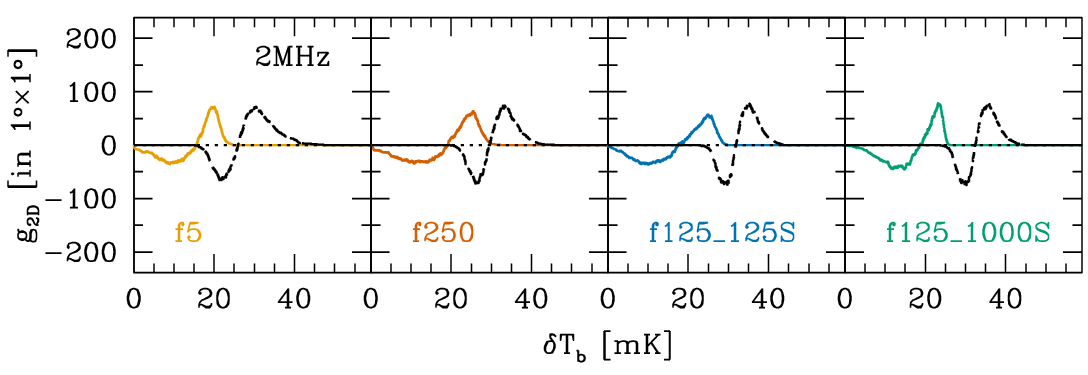}
\caption{2D genus curves of the differential brightness temperatures
  at $x_v = 0.4$, filtered
with a $1\arcmin$ FWHM Gaussian beam and with $0.2 \MHz$ (top), $1 \MHz$
(middle), $2 \MHz$ (bottom) bandwidth;
the curves are compared with those of the matter densities (black).}\label{fig:genus_0_4}
\end{figure*}

\begin{figure*}[!t]
\centering
\includegraphics[width=0.9\textwidth]{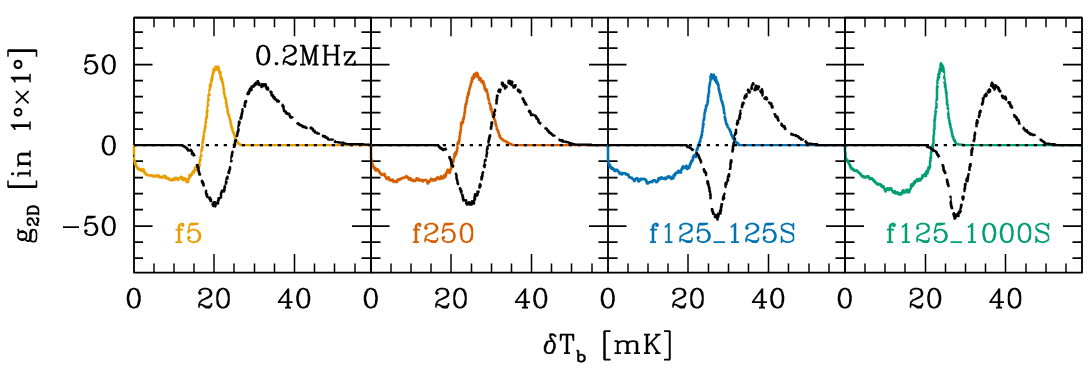}
\includegraphics[width=0.9\textwidth]{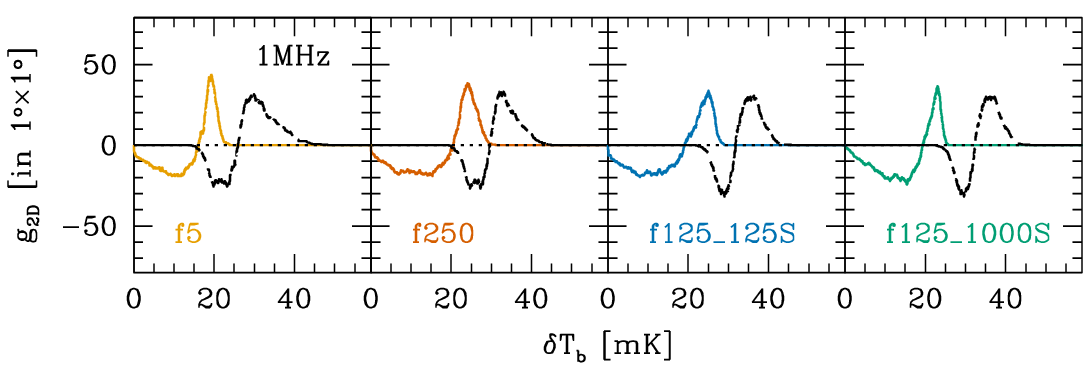}
\includegraphics[width=0.9\textwidth]{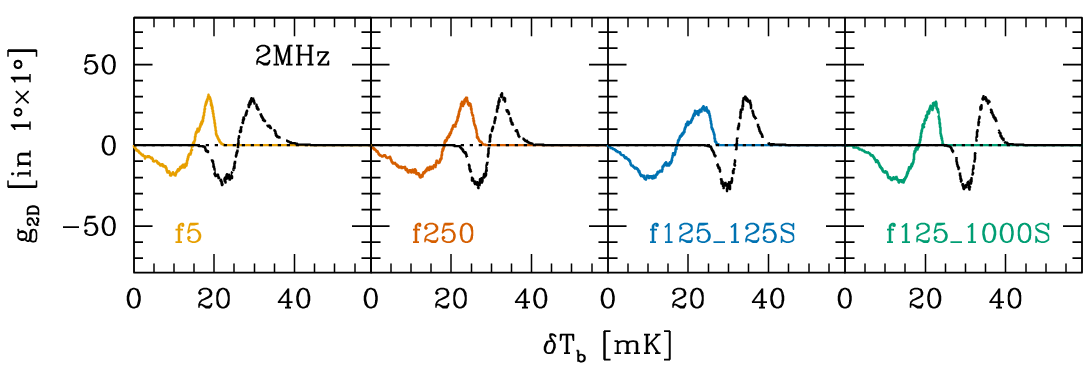}
\caption{Same as Figure~\ref{fig:genus_0_4}, but with a $2\arcmin$ FWHM
  Gaussian beam applied.}\label{fig:genus_0_4_2min}
\end{figure*}

\begin{figure*}[!t]
\centering
\includegraphics[width=0.9\textwidth]{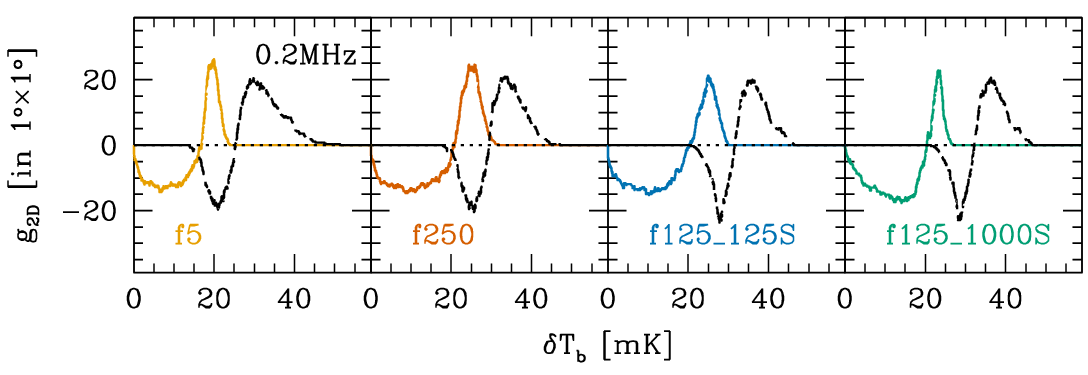}
\includegraphics[width=0.9\textwidth]{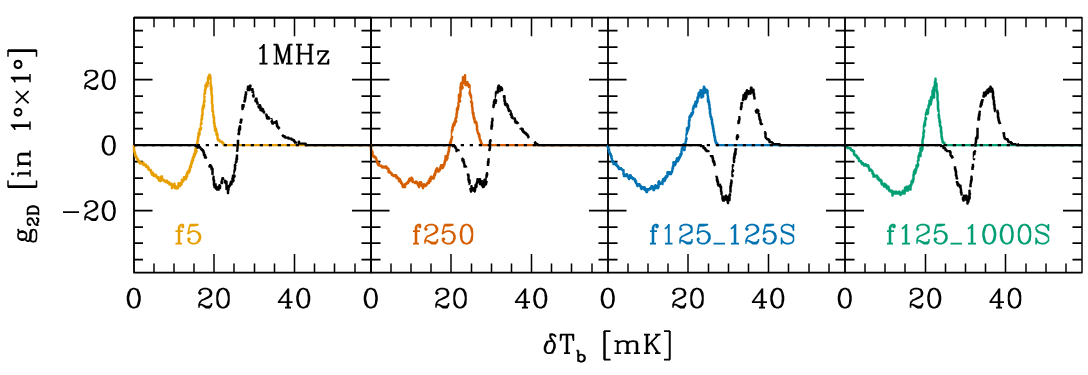}
\includegraphics[width=0.9\textwidth]{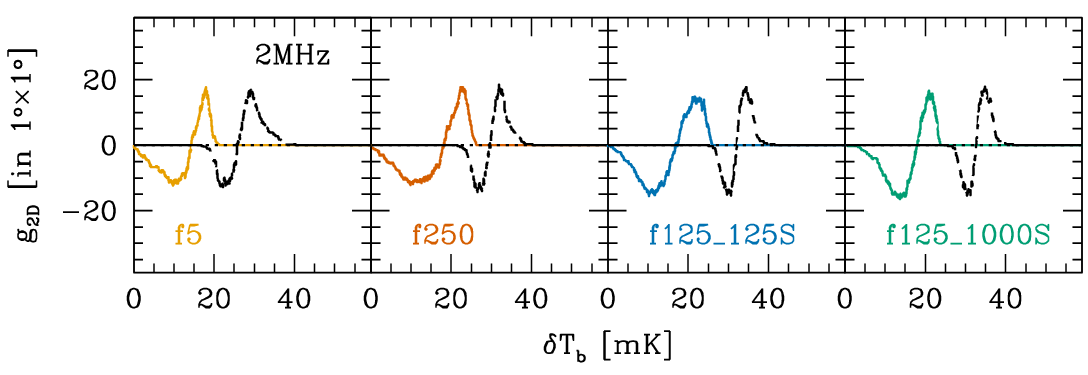}
\caption{Same as Figure~\ref{fig:genus_0_4}, but with a $3\arcmin$ FWHM
  Gaussian beam applied.}\label{fig:genus_0_4_3min}
\end{figure*}

\begin{figure*}[!t]
\begin{center}$
\begin{array}{lcccc}
\textrm{\Large f125\_1000S} & \includegraphics[width=0.2\textwidth]{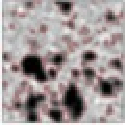} & \includegraphics[width=0.2\textwidth]{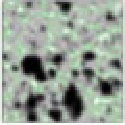} & \includegraphics[width=0.2\textwidth]{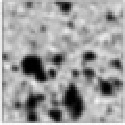} & \includegraphics[width=0.2\textwidth]{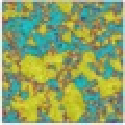} \\
\textrm{\Large f125\_125S} & \includegraphics[width=0.2\textwidth]{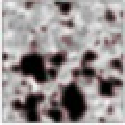} & \includegraphics[width=0.2\textwidth]{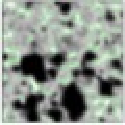} & \includegraphics[width=0.2\textwidth]{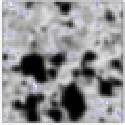} & \includegraphics[width=0.2\textwidth]{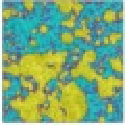} \\[2ex]
& \textrm{\Large a} & \textrm{\Large b} & \textrm{\Large c} & \textit{\Large x} \\
\end{array}$
\end{center}
\caption{How the 2D genus depends on reionization scenarios: from left to
  right, threshold $\dtb$'s (contour lines) are selected, in order to obtain the minimum $g_{\rm 2D}(\dtb)$,
  maximum $g_{\rm 2D}(\dtb)$, and  zero $g_{\rm 2D}(\dtbn)$,
  respectively, along with the snapshot of the unfiltered ionized
  fraction (far right). $x_v = 0.4$ in both (top: f125\_1000S; bottom:
  f125\_125S) cases. See the arrows in Figure~\ref{fig:genus_0_4} for
  the contours a, b and c.}
\label{fig:contour}
\end{figure*}

Our 2D genus analysis seems very promising, indeed, in probing source
properties. The top panel in Figure~\ref{fig:genus_0_4} shows $g_{\rm 2D} (\dtb)$ and
$g_{\rm 2D} (\dtbn)$ for all cases at $x_{\rm v} =
0.4$. The impact of small-mass ($10^8 \lesssim M/M_\odot \lesssim
10^9$) halos is observed as follows. In f125\_1000S, almost all
the (filtered) overdense regions have been ionized at this epoch due to
efficient, small-mass halos. In contrast, when there are no small-mass
halos (f5, f250) or if small-mass halos are not as efficient
(f125\_125S), some fraction of overdense regions still remains neutral
at $x_{\rm v} \sim 0.4$. These regions are mildly nonlinear, and are
ionized due to the almost on-the-spot existence of efficient small-mass
halos in  f125\_1000S, while they mostly remain neutral or only
partially ionized in f5, f250 and f125\_125S due to the absence or
low-efficiency of small-mass halos (Figure~\ref{fig:contour}). The relatively larger amplitude of
$g_{\rm 2D}$ in f125\_1000S is the reflection that there
are many more small individual islands and bubbles, as explained in
Section~\ref{sec:res_genus_evol}.

The impact of beam size and bandwidth is shown in Figures
\ref{fig:genus_0_4}--\ref{fig:genus_0_4_3min}. As
the applied
$\Delta \nu$ increases, the $g_{\rm 2D}$ curve shrinks in width and moves
toward the left. The interpretation we made on f125\_1000S with $\Delta
\nu = 0.2\, {\rm MHz}$ and $\Delta \theta = 1'$
somewhat
weakens, in the sense that $g_{\rm 2D}$ curves are mapped too deep in
the (newly filtered) underdense regions. Nevertheless, the amplitude of $g_{\rm
  2D}$  is the largest and the rightmost wing of $g_{\rm 2D}$ is
located at the smallest $\delta T_b$ in f125\_1000S in all the varying
filtering scales, just as when
$\Delta \nu = 0.2 ' {\rm MHz}$ is applied.

\begin{figure*}[!tp]
\begin{center}$
\begin{array}{ccc}
\includegraphics[width=0.29\textwidth]{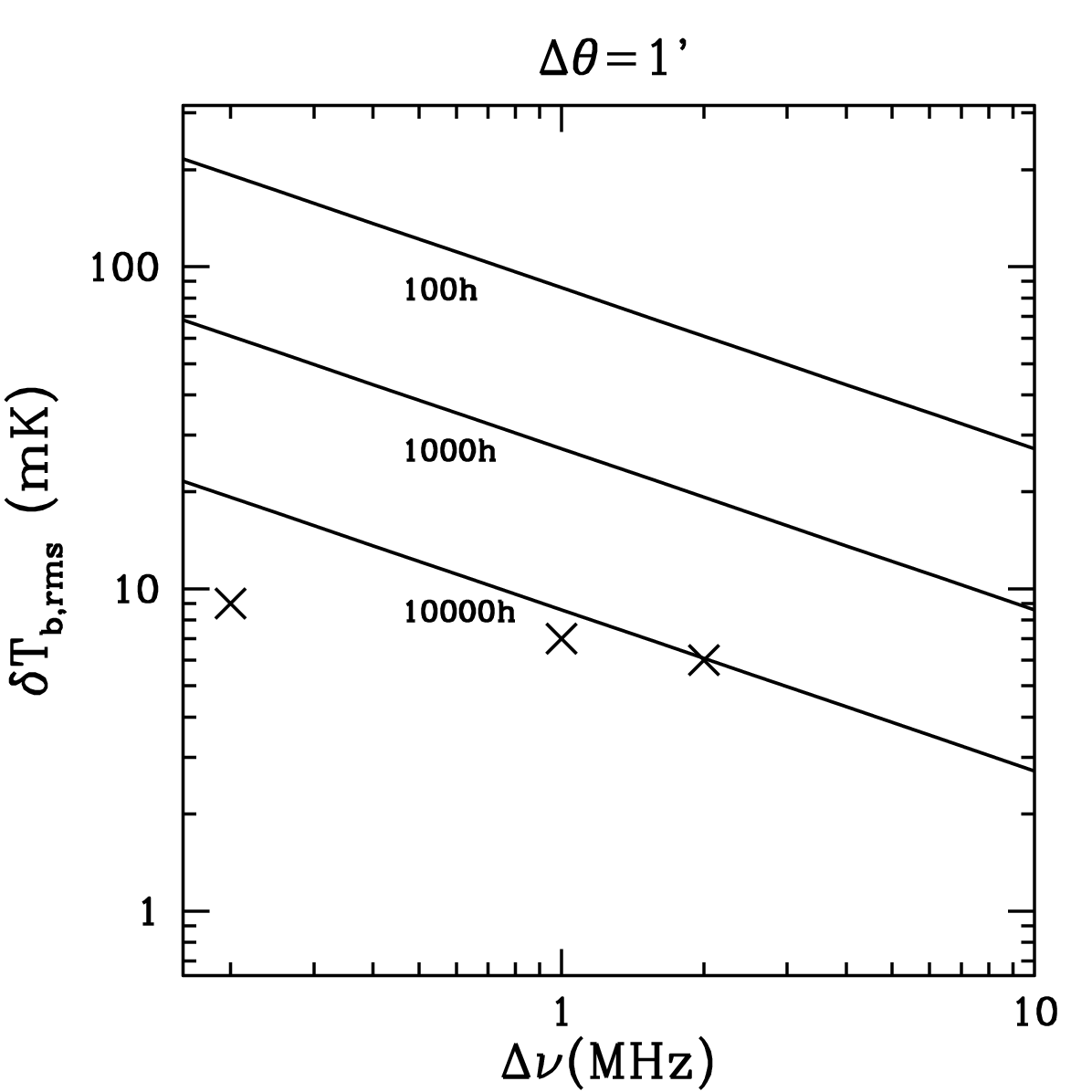} &
\includegraphics[width=0.29\textwidth]{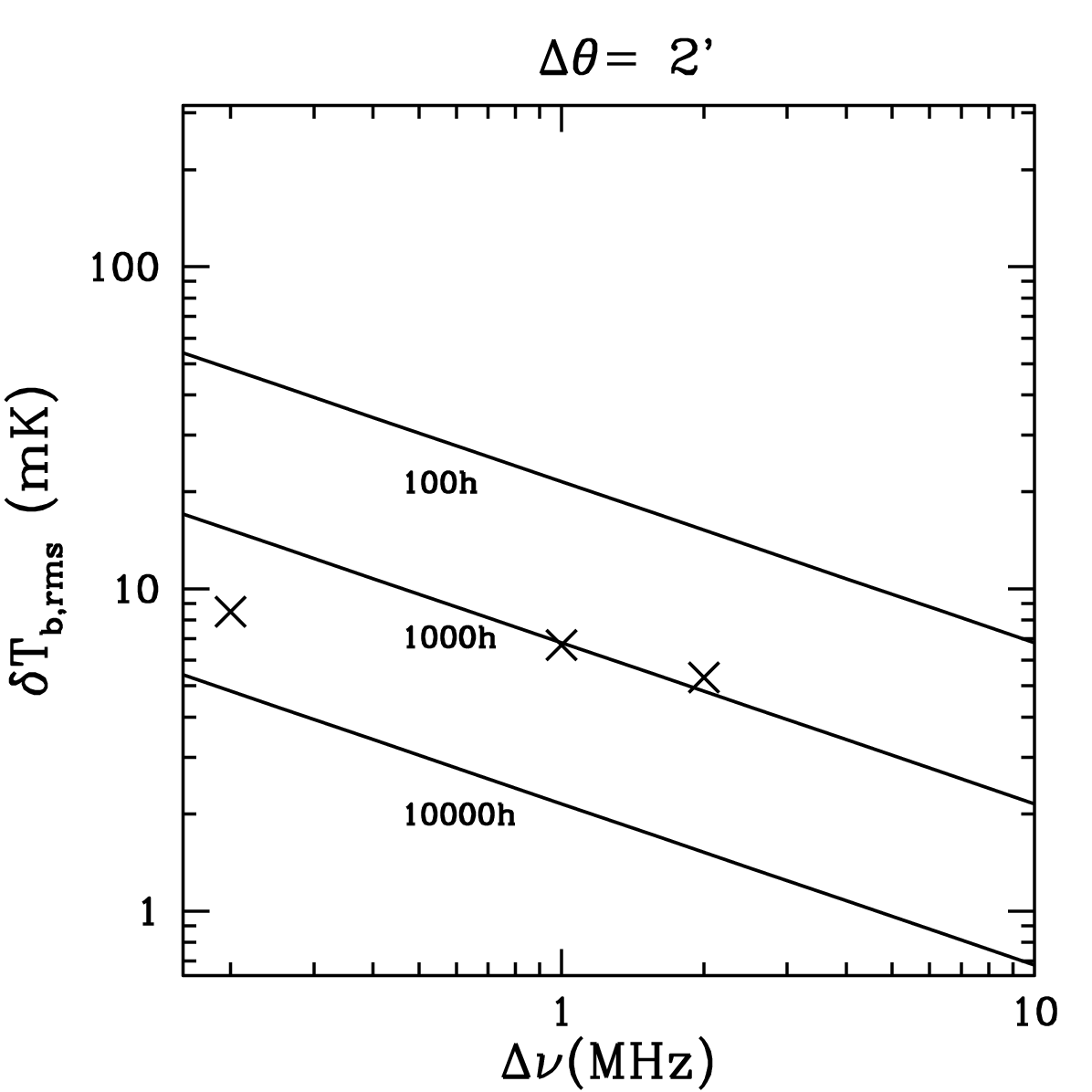} &
\includegraphics[width=0.29\textwidth]{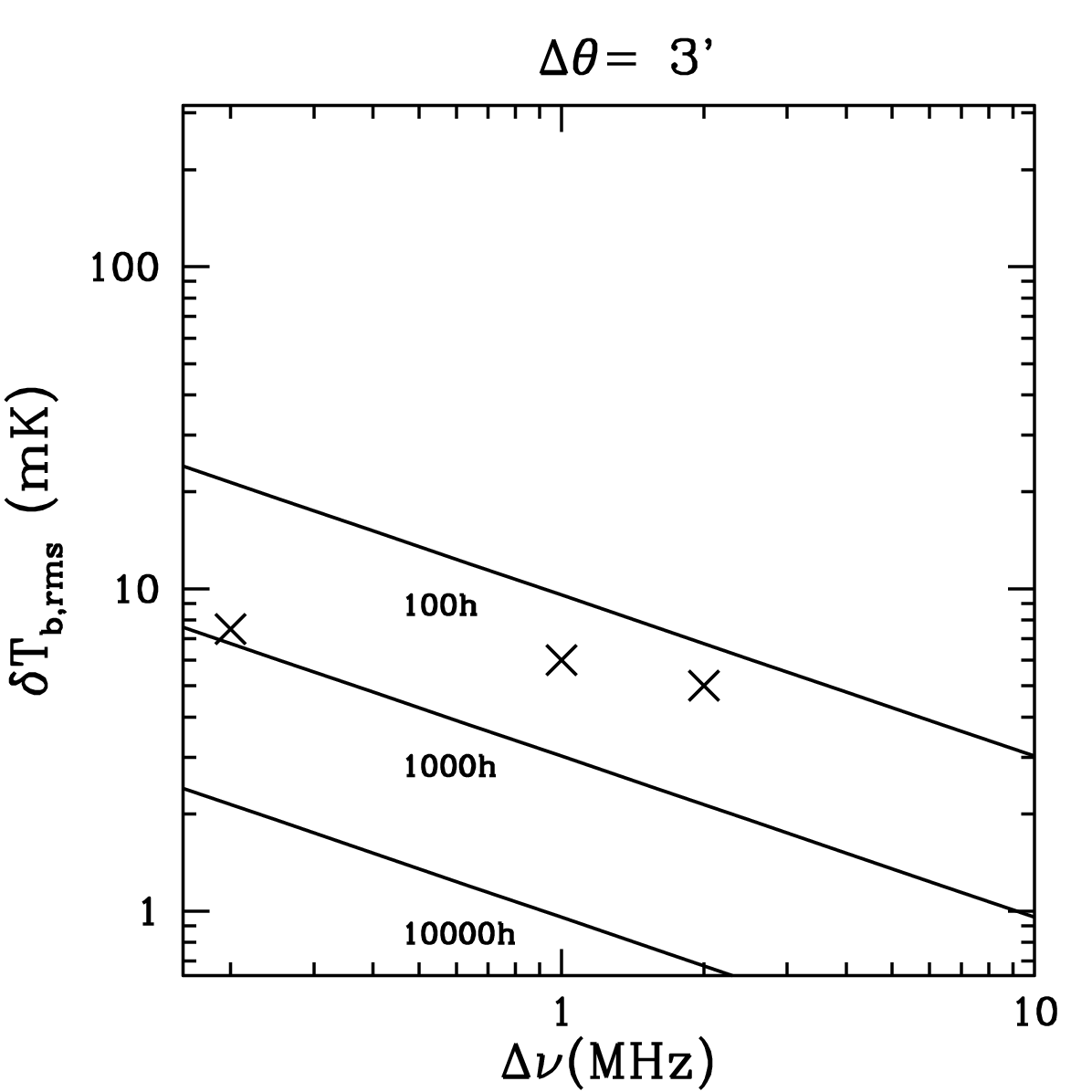} \\
\includegraphics[width=0.3\textwidth]{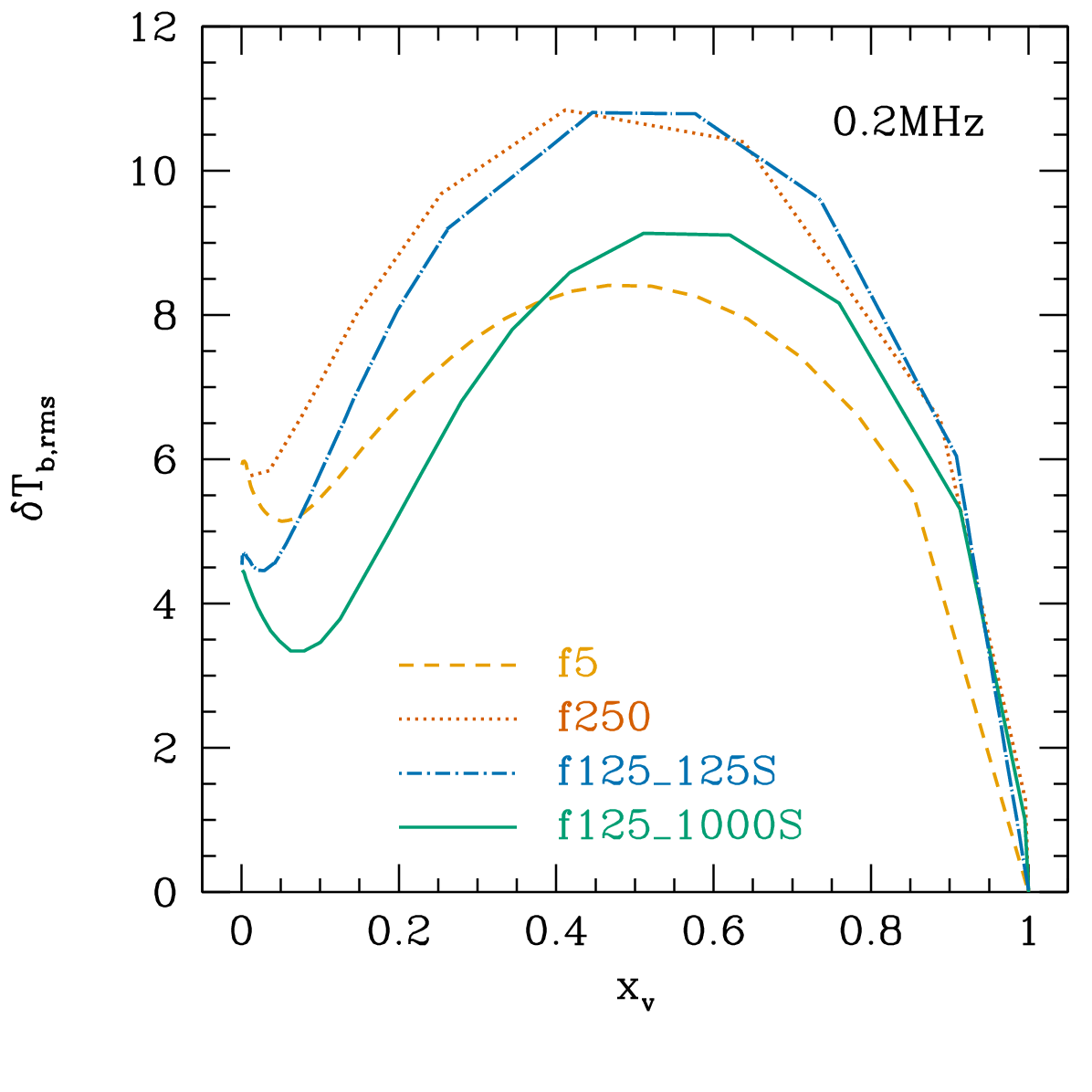} &
\includegraphics[width=0.3\textwidth]{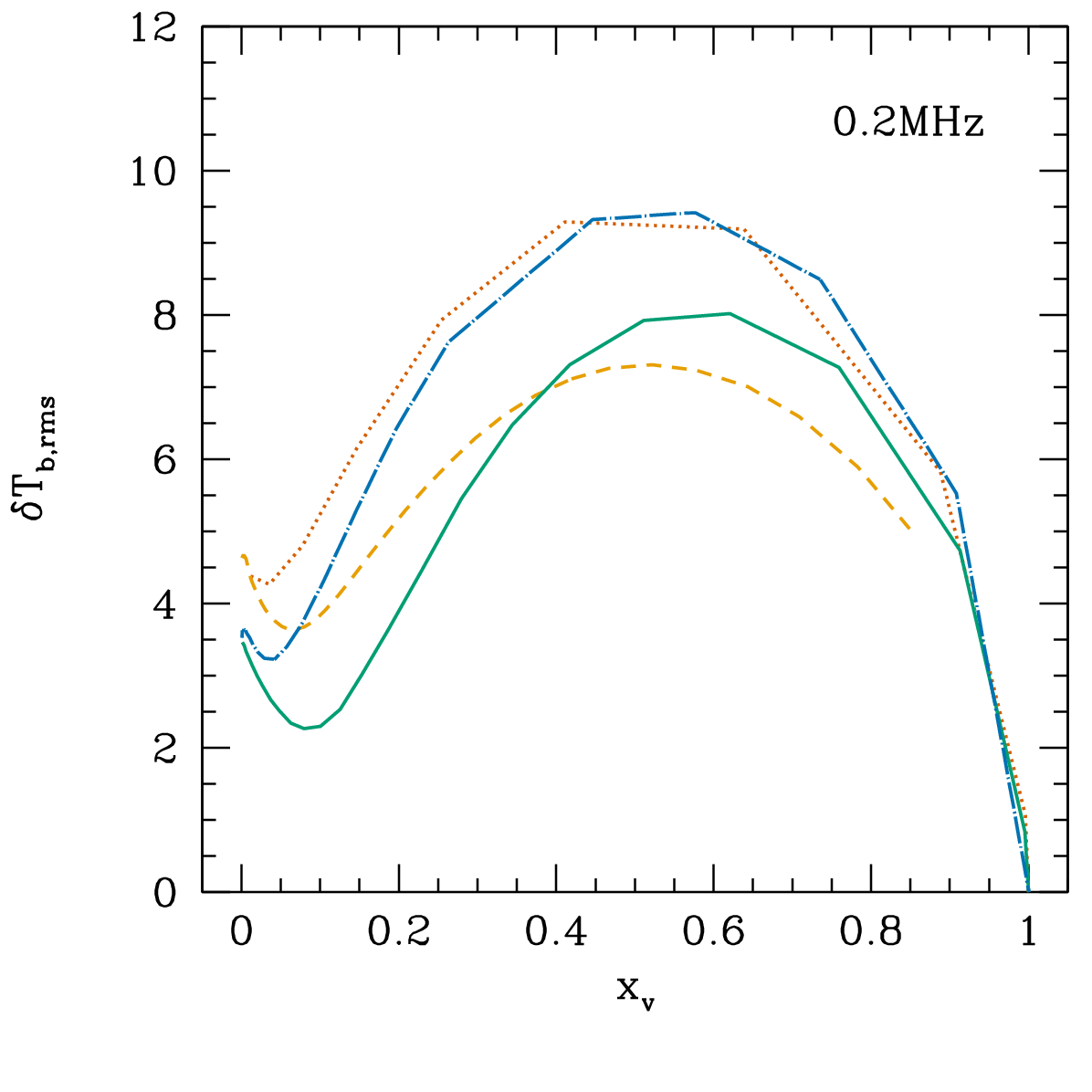} &
\includegraphics[width=0.3\textwidth]{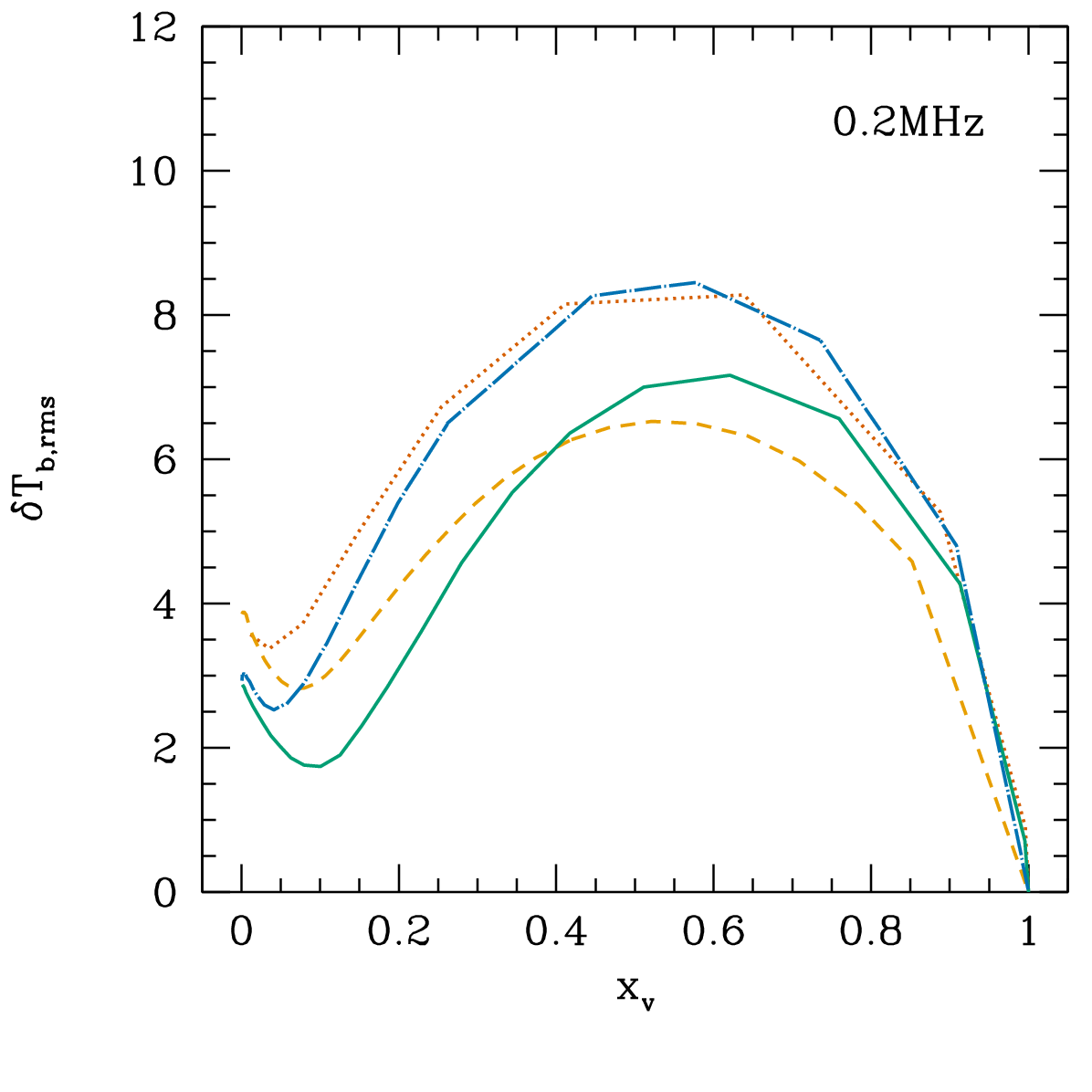} \\
\includegraphics[width=0.3\textwidth]{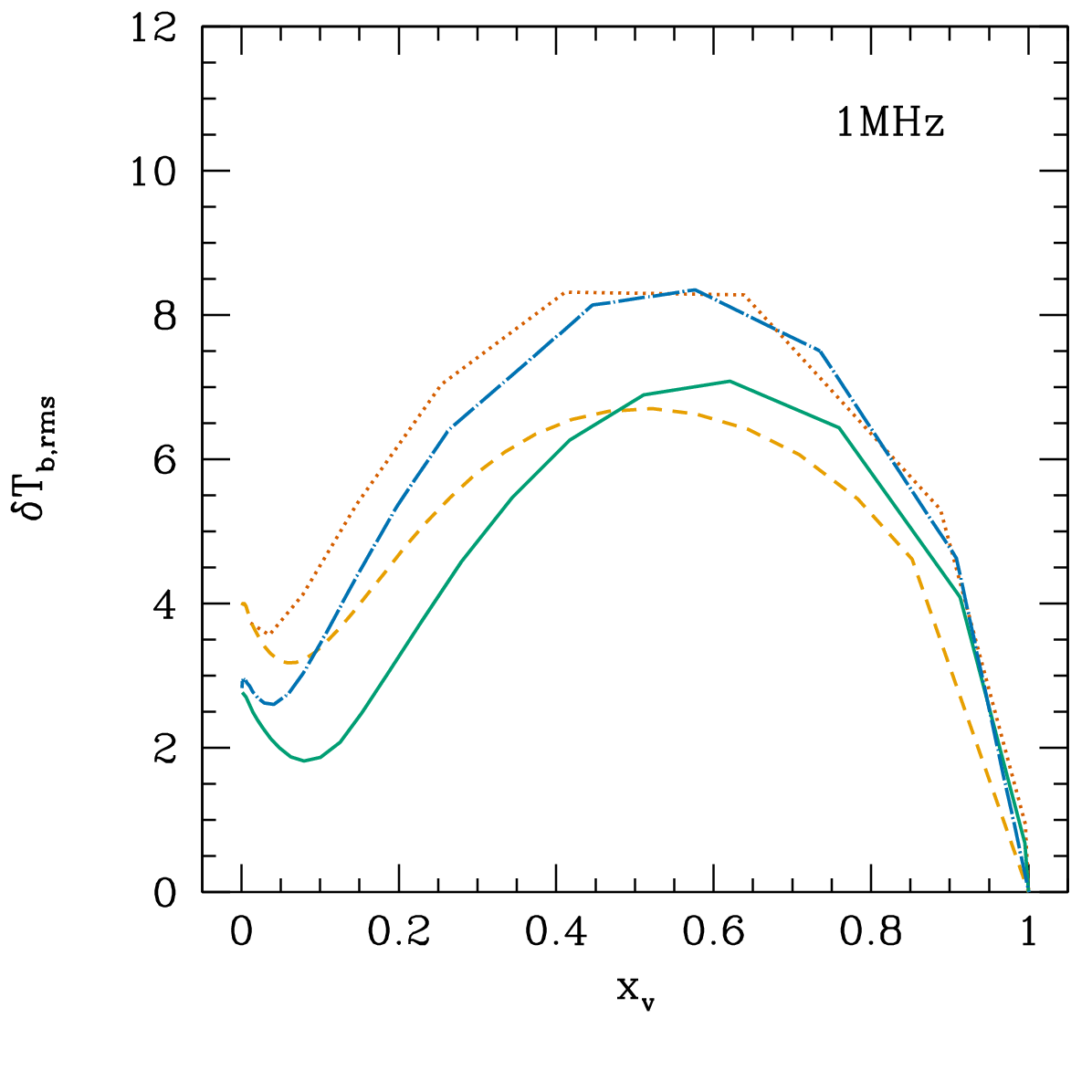} &
\includegraphics[width=0.3\textwidth]{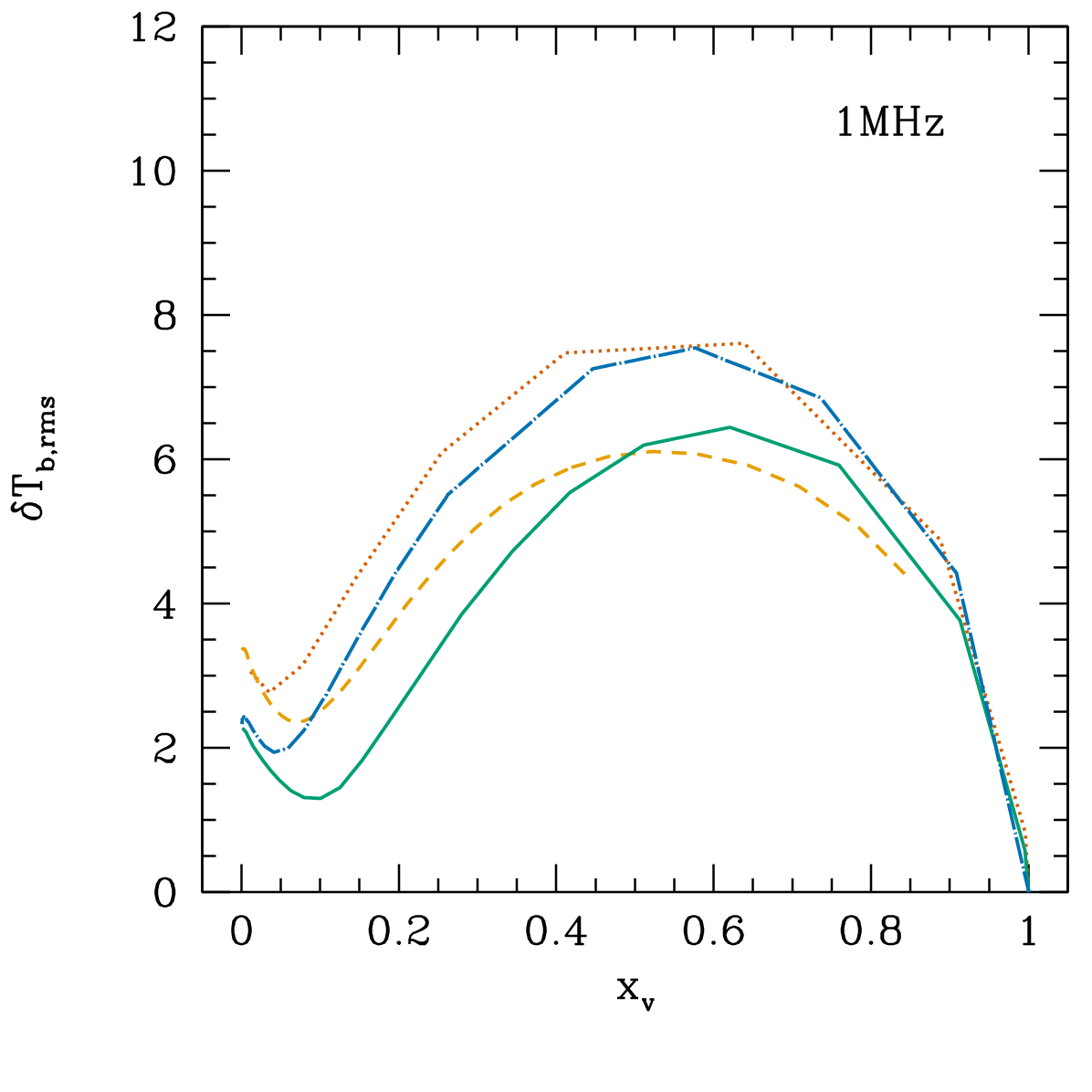} &
\includegraphics[width=0.3\textwidth]{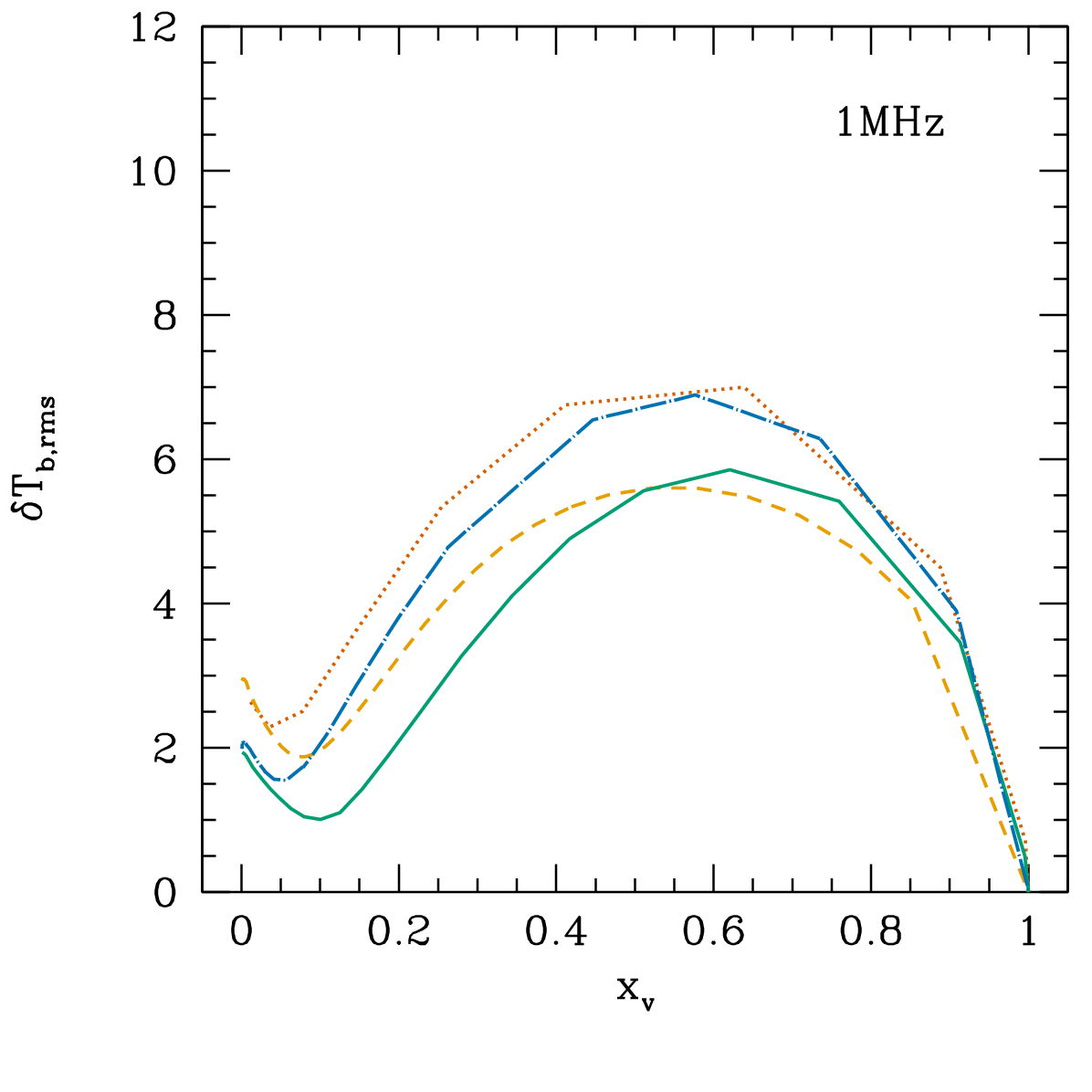} \\
\includegraphics[width=0.3\textwidth]{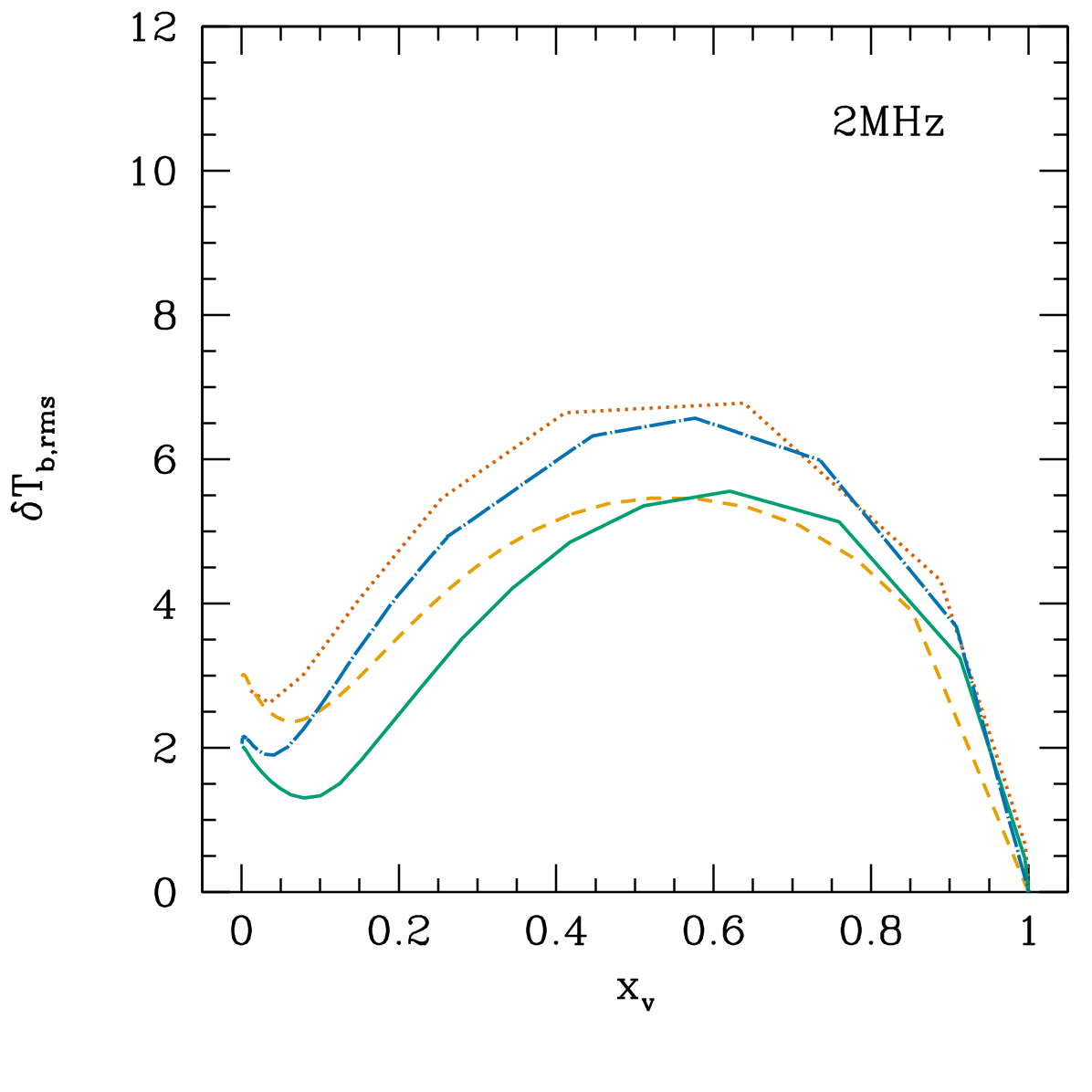} &
\includegraphics[width=0.3\textwidth]{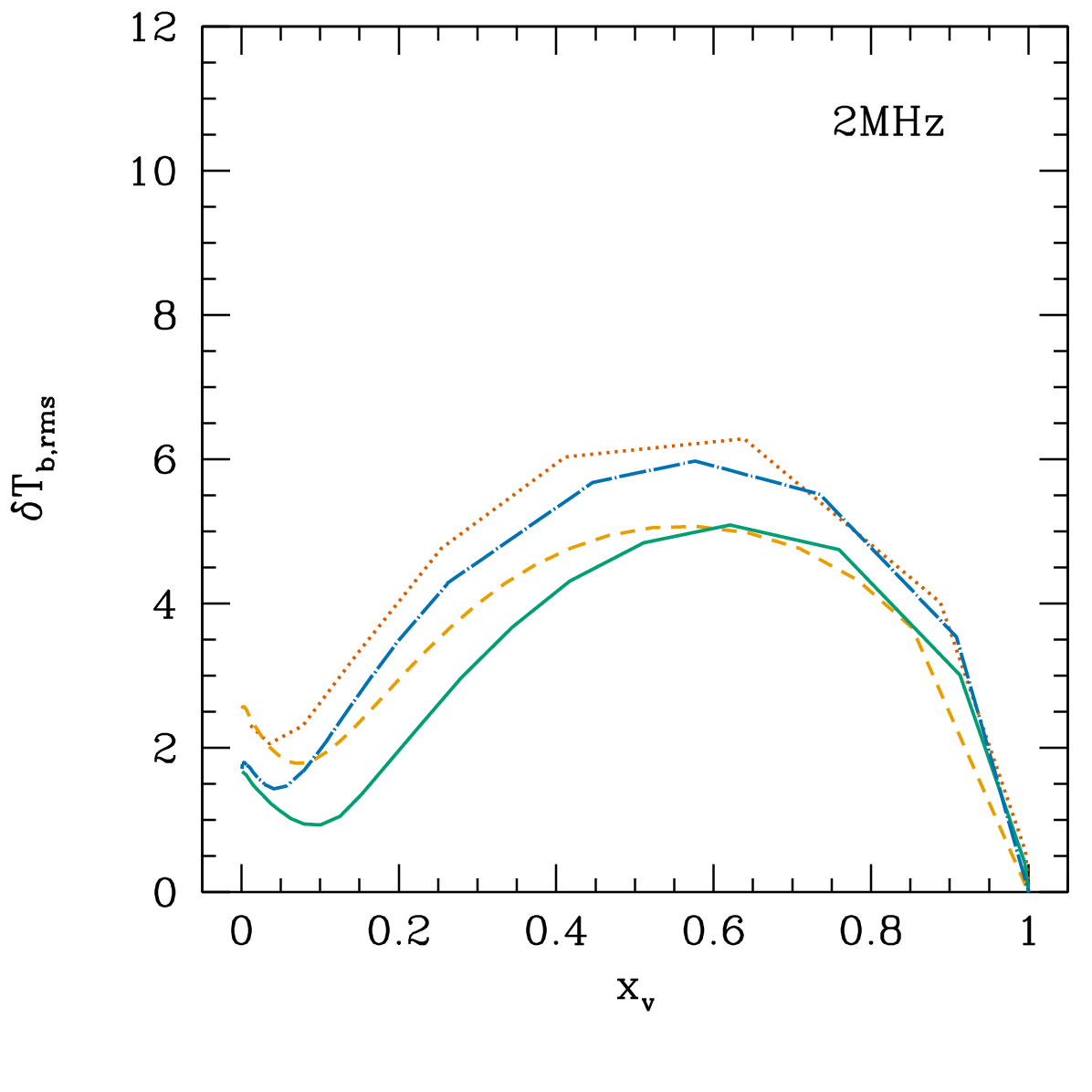} &
\includegraphics[width=0.3\textwidth]{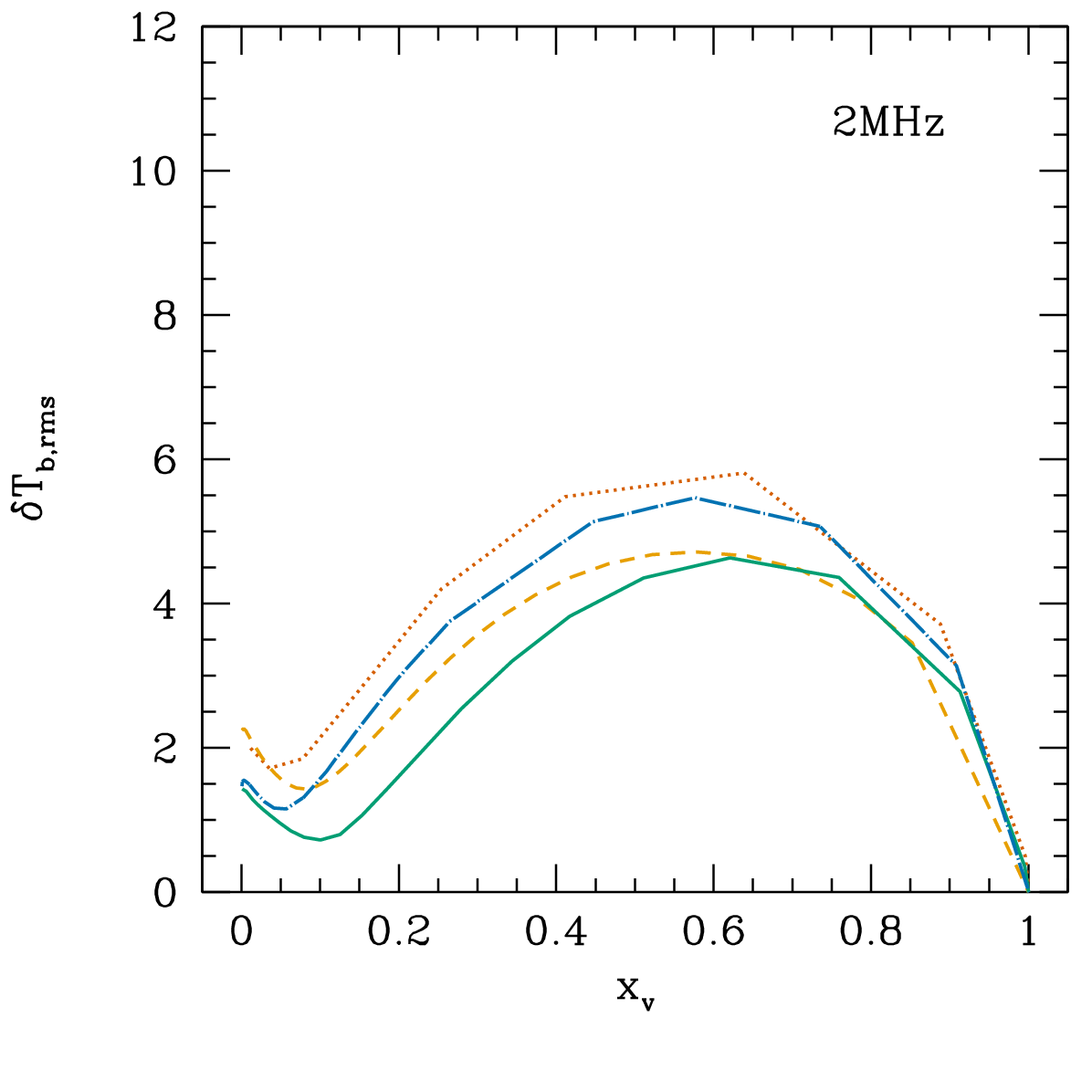} \\
\end{array}$
\caption{Top row: sensitivity limit of SKA in terms of bandwidth
  $\Delta \nu$, with varying integration
  time (100, 1000 and 10000 hours). Also plotted are the required
  sensitivity limits (cross) for useful results, at $\Delta \nu = 0.2$, 1 and
  2 MHz. Each column displays a different beam size $\Delta \theta$'s
  (top labels). Remaining rows: evolution of the rms differential brightness
  temperature as a function of $\Delta \theta$ (each column) and $\Delta
  \nu$ (each row).}
\label{fig:sensitivity}
\end{center}
\end{figure*}

Since $g_{\rm 2D}$ is a useful tool for understanding the
evolution (Section~\ref{sec:res_genus_evol}) and even for discriminating
between reionization
scenarios, we need to ask whether this can be achieved in real
observations. To estimate the required sensitivity, we show the degree
of fluctuation in filtered $\delta T_b$ in Figure
\ref{fig:sensitivity}.
$\delta T_{b,\,\rm rms}$ becomes maximum around the middle stage of
the reionization in all the cases. Roughly speaking, $\delta
T_{b,\,\rm rms}$ gives the required rms sensitivity limit of a
radio antenna in each configuration, which is around a few mK.

Following \citet{Iliev:2003}, we
estimate the sensitivity limit of SKA for the filtering
scale treated in this paper, to figure out the feasibility of our analysis
on real 21-cm background data.
We assume the core size to be $\sim 1\,{\rm km}$
for SKA. We also assume, just for the sake of sensitivity
estimation, that $x_{\rm v} \sim 0.5$ is reached at $z\sim 8$. Even though
our choice of input parameters (properties of the sources) made
reionization ending much earlier
than the standard end of reionization epoch ($z\sim 6.5$),
except for f5, we may imagine that similar distinctive features
among different models would still exist if we tuned these parameters to make
these models achieve the half-ionized state at $z\sim 8$. Figure \ref{fig:sensitivity} shows
the estimated sensitivity under these conditions. Due to the strong
dependence of the sensitivity on $\theta$, $\sim \Delta \theta^{-2}$,
an increase in the beam size quickly makes our analysis feasible. This
comes at a price that values of $g_{\rm 2D}(\delta T_b )$ shrink
further, and the range of $\delta T_b$ shifts more to the left.
At any rate, $\sim 1000 -
10000$ hours of integration at $\delta \nu \sim 1 - 2 $ MHz and
$\delta \theta \sim 2' - 3'$ seem possible to allow for our analysis.

\section{Summary and discussion}\label{sec:concl}

We calculated the 21-cm radiation background from high redshift using
a suite of structure formation and radiative transfer simulations with
different properties of the sources of radiation. These properties
are specified by the spectrum of halo masses capable of
hosting the sources of radiation and the emissivity of hydrogen-ionizing
photons. Assuming the pre-reionization heating of the IGM, we calculated
the 21-cm radiation background in the saturated limit, $T_{\rm S} \gg
T_{\rm CMB}$, such that its differential brightness temperature is
proportional to the underlying density and neutral fraction.

In order to understand the topology of the high-redshift IGM, we
developed 2D genus method applicable to the 21-cm
radiation background.
Basically, this method calculates the 2D genus,
the difference between the number of hot and cold spots, under
a given threshold differential brightness temperature. We construct 2D
genus curves at different redshifts for different scenarios, by
varying this threshold values.

We found that the 2D genus curve $\fn{g_{\rm 2D}}{\dtb}$, if
compared to the $g_{\rm 2D}$ of the underlying density fluctuation $\fn{g_{\rm 2D}}{\dtbn}$,
clearly shows the evolution of the reionization
process. The $\fn{g_{\rm 2D}}{\dtb}$ curve is found to deviate quickly from
the $g_{\rm 2D}$ of a Gaussian random field, qualitatively in agreement with
the findings of \citet{Iliev:2006} and in contrast with the
findings of \citet{Shin:2008}. \citet{Shin:2008} find that the
non-Gaussianity of the H II region is small, even when the Universe is
50\% ionized.
It is also shown that the reionization
proceeds in an inside-out fashion: high-density regions get ionized first
and low-density regions later.

We also showed that our 2D genus method can be used to
discriminate between various reionization scenarios, thus probing
the properties of the sources indirectly.
It appears to be most effective in discriminating the mass spectra of halos
which host the sources of radiation. A hybrid mass spectrum with a different
emissivity in different species (f125\_1000S in this paper) stands out
from the cases with a single species (f5, f250) or a case with
equal-emissivity among different species (f125\_125S).

Crucial ingredients needed for this analysis are beam shapes and
the sensitivity. We tested two different beams, a Gaussian and a compensated
Gaussian. The Gaussian beam, even after the field is filtered and
degraded, leaves a distinguishable imprint on the 2D genus curve such that
different reionization scenarios can be discriminated. In contrast,
the compensated Gaussian beam somewhat loses the interpretative power
that the Gaussian beam had.
Assuming that the compensated Gaussian beam represents the so-called
dirty beams, our analysis
may be applied entirely only when those artificial
effects from dirty beams are removed.

We predict that SKA will be able to produce data suitable for this
analysis, when $1000 - 10000$ hours of integration are performed
with $\delta \nu \sim 2 - 3\, {\rm MHz}$ and $\delta \theta \sim 2' -
3'$ at an observing frequency of $\sim 150\,{\rm MHz}$. Therefore,
our method seems promising.
Note, however, that a direct link
from the observed 2D genus curve to the true properties of the sources will
be still far-fetched, when we adopt such low-resolution filters to
increase the sensitivity.
Moreover, there are still too many uncertainties, such as
the
matter power spectrum at small scales, which is relevant to the formation
and evolution of the sources of reionization. We hope to see many more
useful constraints on cosmic reionization from other various,
direct or indirect,
observations. In the future, we will explore more reionization
scenarios to strengthen the potential of our 2D genus analysis.

One caveat of our work is that the simulation box is relatively small,
while a $\gtrsim 100$ Mpc box is likely to provide a more statistically
reliable result. While 2D genus should be dominated by the most
abundant H II bubbles, which are also the smallest bubbles, the cosmic
variance guarantees that some bubble sizes, at late stages of reionization,
will become larger than the box we used in this work. We have indeed
simulated reionization in a very large
box ($425/h$ Mpc) and observed a variation of the observational
properties (e.g. 3D power spectrum) in a certain range of scales and
reionization stages \citep{Iliev:2013}. We did not yet analyze the minihalo-included
simulation \citep{Ahn:2012}, which forms another important class of
models with very small H II regions. We
thus plan to apply our 2D genus analysis to these new results (more
model dependency and bigger box size) and
provide a more statistically reliable result which can be used
to analyze future 21-cm observations.

\acknowledgments

KA was supported in part by NRF grant funded by the Korean government MEST (No. 2012R1A1A1014646),
and KA also acknowleges the generous support from Chosun University for KA's sabbatical (research) year.
We thank Korea Institute for Advanced Study for providing computing resources (KIAS Center for Advanced Computation Linux Cluster System) for this work.
ITI was supported by the Science and
Technology Facilities Council [grant numbers ST/F002858/1 and ST/I000976/1];
and The Southeast Physics Network (SEPNet).
GM was supported by Swedish Research Council grant 2012-4144.


\begin{thebibliography}{}

\bibitem[Adler (1981)]{Adler:1981}
  Adler, R.~J.\ 1981,
  The Geometry of Random Fields, Wiley, New York

\bibitem[Ahn \etal(2009)]{Ahn:2008sq}
  Ahn, K., Shapiro, P.~R., Iliev, I.~T., Mellema, G., \& Pen, U.~L.\ 2009,
  The Inhomogeneous Background of H2 Dissociating Radiation During Cosmic
  Reionization,
  \apj, 695, 1430

\bibitem[Ahn \etal(2012)]{Ahn:2012}
  Ahn, K., Iliev, I.~T., Shapiro, P.~R., Mellema, G., Koda, J., \& Mao, Y.\ 2012,
  Detecting the Rise and Fall of the First Stars by Their Impact on Cosmic Reionization,
  \apj, 756, L16
	
\bibitem[Cen (2003)]{Cen:2003ey}
  Cen, R.\ 2003,
  Implications of WMAP Observations On the Population III Star Formation
  Processes,
  \apj, 591, L5

\bibitem[Chepurnov \etal(2008)]{Chepurnov:2008}
  Chepurnov, A., Gordon, J., Lazarian, A., \& Stanimirovic, S.\ 2008,
	Topology of Neutral Hydrogen within the Small Magellanic Cloud,
  \apj, 688, 1021

\bibitem[Ciardi \& Madau(2003)]{Ciardi:2003hg}
  Ciardi, B., \& Madau, P.\ 2003,
  Probing Beyond the Epoch of Hydrogen Reionization with 21 Centimeter
  Radiation,
  \apj, 596, 1

\bibitem[Coles(1989)]{Coles:1989su}
  Coles, P.\ 1989,
  The Clustering of Local Maxima in Random Noise,
  \mnras, 238, 319

\bibitem[Colley(1997)]{Colley:1996gh}
  Colley, W.~N.\ 1997,
  Two Dimensional Topology of Large Scale Structure in the Las Campanas
  Redshift Survey,
  \apj, 489, 471

\bibitem[Colley, Gott, \& Park(1996)]{Colley:1996vj}
  Colley, W.~N., Gott, J.~R., \& Park, C.\ 1996,
  Topology of COBE Microwave Background Fluctuations,
  \mnras, 281, L82

\bibitem[Colley \& Gott(2003)]{Colley:2003sp}
  Colley, W.~N., \& Gott, J.~R.~I.\ 2003,
  Genus Topology of the Cosmic Microwave Background from WMAP,
  \mnras, 344, 686

\bibitem[Datta \etal(2012)]{Datta:2012} Datta, K.~K.,
  Mellema, G., Mao, Y., et al.\ 2012,
  Light-Cone Effect on the Reionization 21-cm Power Spectrum,
  \mnras, 424, 1877


\bibitem[Dijkstra \etal(2008)]{Dijkstra:2008jk}
  Dijkstra, M., Haiman, Z., Mesinger, A., \& Wyithe, S.\ 2008,
  Fluctuations in the High-Redshift Lyman-Werner Background: Close Halo Pairs
  as the Origin of Supermassive Black Holes,
  \mnras, 391, 1961

\bibitem[Dubinski \etal(2004)]{Dubinski:2003fq}
  Dubinski, J., Kim. J., Park, C., \& Humble, R.\ 2004,
  GOTPM: A Parallel Hybrid Particle-Mesh Treecode,
  New Astron., 9, 111

\bibitem[Dunkley \etal(2009)]{Dunkley:2008ie}
	Dunkley, J., Spergel, D.~N., Komatsu, E., Hinshaw, G., Larson, D.,
	Nolta, M.~R., Odegard, N., Page, L., Bennett, C.~L., Gold, B.,
	Hill, R.~S., Jarosik, N., Weiland, J.~L., Halpern, M., Kogut, A.,
	Limon, M., Meyer, S.~S., Tucker, G.~S., Wollack, E., \& Wright, E.~L.\ 2009,
  Five-Year Wilkinson Microwave Anisotropy Probe (WMAP) Observations:
  Likelihoods and Parameters from the WMAP data,
  \apjs, 180, 306

\bibitem[Efstathiou \etal(1985)]{Efstathiou:1985re}
  Efstathiou, G., Davis, M., Frenk, C.~S., \& White, S.~D.~M.\ 1985,
  Numerical Techniques For Large Cosmological N-Body Simulations,
  \apjs, 57, 241

\bibitem[Fan \etal(2006)]{Fan:2005es}
	Fan, X., Strauss, M.~A., Becker, R.~H., White, R.~L., Gunn, J.~E.,
	Knapp, G.~R., Richards, G.~T., Schneider, D.~P., Brinkmann, J.,
	\& Fukugita, M.\ 2006,
  Constraining the Evolution of the Ionizing Background and the Epoch of
  Reionization with z $\sim$ 6 Quasars II: A Sample of 19 Quasars,
  \aj, 132, 117

\bibitem[Field(1959)]{Field:1959}
	Field, G.~B.\ 1959,
	The Spin Temperature of Intergalactic Neutral Hydrogen,
	\apj, 129, 536
	
\bibitem[Friedrich \etal(2011)]{Friedrich:2010}
	Friedrich, M.~M., Mellema, G., Alvarez, M.~A., Shapiro, P.~R., \& Iliev, I.~T.\ 2011,
	Topology and Sizes of H II Regions during Cosmic Reionization,
	\mnras, 413, 1353
	
\bibitem[Gleser \etal(2006)]{Gleser:2006}
  Gleser, L., Nusser, A., Benedetta, C., \& Desjacques, V.\ 2006,
	The Morphology of Cosmological Reionization by means of Minkowski Functionals,
  \mnras, 370, 1329

\bibitem[Gott \etal(1990)]{Gott:1989yj}
  Gott, J.~R., Park, C., Juszkiewicz, R., Bies, W.~E., Bennett, D.~P., Bouchet, F.~R., \& Stebbins, A.\ 1990,
  Topology of Microwave Background Fluctuations: Theory,
  \apj, 352, 1

\bibitem[Gott \etal(1992)]{Gott:1992}
  Gott. J.~R., Mao. S., Park, C., \& Lahav. O.\ 1992,
  The Topology of Large-Scale Structure. V. - Two Dimensional Topology of Sky Map,
  \apj, 385, 26

\bibitem[Gott \etal(2007)]{Gott:2006za}
  Gott, J.~R., Colley, W.~N., Park, C.~G., Park, C., \& Mugnolo, C.\ 2007,
  Genus Topology of the Cosmic Microwave Background from the WMAP 3-Year
  Data,
  \mnras, 377, 1668

\bibitem[Gott \etal(2009)]{Gott:2008kk}
  Gott, J.~R., Choi, Y.~Y.~I., Park, C., \& Kim, J.\ 2009,
  3D Genus Topology of Luminous Red Galaxies,
  \apj, 695, L45

\bibitem[Gott, Melott \& Dickinson(1986)]{Gott:1986uz}
  Gott. J.~R., Melott, A.~L., \& Dickinson, M.\ 1986,
  The Sponge-like Topology of Large-Scale Structure in the Universe,
  \apj, 306, 341

\bibitem[Gott, Weinberg \& Melott(1987)]{Gott:1987}
  Gott, J.~R., Weinberg, D.~H., \& Melott, A.~L.\ 1987,
  A Quantitative Approach to the Topology of Large-Scale Structure,
  \apj, 319, 1

\bibitem[Haiman, Abel \& Rees(2000)]{Haiman:1999mn}
  Haiman, Z., Abel, T., \& Rees, M.~J.\ 2000,
  The Radiative Feedback of the First Cosmological Objects,
  \apj, 534, 11

\bibitem[Haiman \& Bryan(2006)]{Haiman:2006si}
  Haiman, Z., \& Bryan, G.~L.\ 2006,
  Was Star-Formation Suppressed in High-Redshift Minihalos?,
  \apj, 650, 7

\bibitem[Hamilton, Gott \& Weinberg(1986)]{Hamilton:1986}
  Hamilton, A.~J.~S., Gott, J.~R., \& Weinberg, D.~H.\ 1986,
	The Topology of the Large-Scale Structure of the Universe,
  \apj, 309, 1

\bibitem[Hoyle \& Vogeley(2002)]{Hoyle:2001kn}
  Hoyle, F., \& Vogeley, M.~S.\ 2002,
  Voids in the PSCz Survey and the Updated Zwicky Catalog,
  \apj, 566, 641

\bibitem[Hoyle, Vogeley \& Gott(2002)]{Hoyle:2001xx}
  Hoyle, F., Vogeley. M.~S., \& Gott, J.~R.\ 2002,
  Two-Dimensional Topology of the 2dF Galaxy Redshift Survey,
  \apj, 570, 44

\bibitem[Hui \& Haiman(2003)]{Hui:2003hn}
  Hui, L., \& Haiman, Z.\ 2003,
  The Thermal Memory of Reionization History,
  \apj, 596, 9

\bibitem[Ikeuchi(1986)]{Ikeuchi:1985rf}
  Ikeuchi, S.\ 1986,
  Baryon Clump within an Extended Dark Matter,
  \apss, 118, 509

\bibitem[Iliev \etal(2002)]{Iliev:2002gj}
  Iliev, I.~T., Shapiro, P.~R., Ferrara, A., \& Martel, H.\ 2002,
  On the Direct Detectability of the Cosmic Dark Ages: 21-cm Emission from
  Minihalos,
  \apj, 572, 123

\bibitem[Iliev \etal(2003)]{Iliev:2003}
  Iliev, I.~T., Scannapieco, E., Martel, H., \& Shapiro, P.~R.\ 2003,
	Non-Linear Clustering during the Cosmic Dark Ages and its Effect
	on the 21-cm Background from Minihaloes,
  \mnras, 341, 81

\bibitem[Iliev \etal(2005)]{Iliev:2005}
	Iliev, I.~T., Shapiro, P.~R., \& Raga, A.~C.\ 2005,
	Minihalo Photoevaporation during Cosmic Reionization: Evaporation Times and Photon Consumption Rates,
	\mnras, 361, 405

\bibitem[Iliev \etal(2006)]{Iliev:2006}
  Iliev, I.~T., Mellema, G., Pen, U.-L., Merz, H., Shapiro, P.~R., \& Alvarez, M.~A.\ 2006,
  Simulating Cosmic Reionization at Large Scales
	-- I. The Geometry of Reionization,
  \mnras, 369, 1625

\bibitem[Iliev \etal(2007)]{Iliev:2006sw}
  Iliev, I.~T., Mellema, G., Shapiro, P.~R., \& Pen, U.~L.\ 2007,
  Self-Regulated Reionization,
  \mnras, 376, 534

\bibitem[Iliev \etal(2013)]{Iliev:2013}
	Iliev, I.~T., Mellema, G., Ahn, K., Shapiro, P.~R., Mao, Y., \& Pen, U.~L.\ 2013,
	Simulating Cosmic Reionization: How Large a Volume is Large Enough?,
	arXiv:1310.7463

\bibitem[Kim \& Park(2007)]{Kim:2007}
  Kim, S., \& Park, C.\ 2007,
	Topology of H I Gas Distribution in the Large Magellanic Cloud,
  \apj, 663, 244

\bibitem[Kim \etal(2009)]{Kim:2008kf}
  Kim, J., Park, C., Gott, J.~R., \& Dubinski, J.\ 2009,
  The Horizon Run N-body Simulation: Baryon Acoustic Oscillations and
  Topology of Large Scale Structure of the Universe,
  \apj, 701, 1547

\bibitem[Kim \& Pen(2010)]{Kim:2010}
	Kim, J., \& Pen, U.-L.\ 2010,
	Redshifted 21-cm Signals in the Dark Ages,
	ArXiv e-prints:0908.1973

\bibitem[Kogut(1993)]{Kogut:1993}
  Kogut, A.\ 1993,
  Topology of the COBE-DMR First Year Sky Map,
  \baas, 183, 121.03

\bibitem[Kogut \etal(1996)]{Kogut:1996uu}
  Kogut, A., Banday, A.~J., Bennett, C.~L., Gorski, K., Hinshaw, G., Smoot, G.~F., \& Wright, E.~L.\ 1996,
  Tests for Non-Gaussian Statistics in the DMR Four-Year Sky Maps,
  \apj, 464, L29

\bibitem[Komatsu et al.(2009)]{Komatsu:2008hk}
	Komatsu, E., Dunkley, J., Nolta, M.~R., Bennett, C.~L., Gold, B.,
	Hinshaw, G., Jarosik, N., Larson, D., Limon, M., Page, L.,
	Spergel, D.~N., Halpern, M., Hill, R.~S., Kogut, A., Meyer, S.~S.,
	Tucker, G.~S., Weiland, J.~L., Wollack, E., \& Wright, E.~L.\ 2009,
  Five-Year Wilkinson Microwave Anisotropy Probe (WMAP)
  Observations: Cosmological Interpretation,
  \apjs,  180, 330

\bibitem[La Plante \etal(2013)]{LaPlante:2013} La Plante, P.,
  Battaglia, N., Natarajan, A., et al.\ 2013,
  Reionization on Large Scales IV: Predictions for the 21 cm Signal Incorporating the Light Cone Effect,
  arXiv:1309.7056

\bibitem[Lee \etal(2008)]{Lee:2008}
  Lee, K.-G., Cen, R., Gott, J.~R., \& Trac, H.\ 2008,
	The Topology of Cosmological Reionization,
  \apj, 675, 8

\bibitem[Madau, Meiksin \& Rees(1997)]{Madau:1996cs}
  Madau, P., Meiksin, A., \& Rees, M.~J.\ 1997,
  21-cm Tomography of the Intergalactic Medium at High Redshift,
  \apj, 475, 429

\bibitem[McQuinn \etal(2007)]{McQuinn:2007}
  McQuinn, M., Lidz, A., Zahn, O., Dutta, S., Hernquist, L., \& Zaldarriaga, M.\ 2007,
  The Morphology of H II Regions during Reionization,
  \mnras, 377, 1043

\bibitem[Mellema \etal(2006a)]{Mellema:2005ht}
  Mellema, G., Iliev, I.~T., Alvarez, M,~A., \& Shapiro, P.~R.\ 2006,
  C$^2$-Ray: A New Method for Photon-Conserving Transport of Ionizing
  Radiation,
  New Astron.\, 11, 374

\bibitem[Mellema \etal(2006b)]{Mellema:2006pd}
  Mellema, G., Iliev, I.~T., Pen, U.~L., \& Shapiro, P.~R.\ 2006,
  Simulating Cosmic Reionization at Large Scales II: the 21-cm Emission
  Features and Statistical Signals,''
  \mnras, 372, 679

\bibitem[Mellema \etal(2013)]{Mellema:2013}
	Mellema, G., Koopmans, L.~V.~E., Abdalla, F.~A., Bernardi, G., Ciardi, B.,
	Daiboo, S., de Bruyn, A.~G., Datta, K.~K., Falcke, H., Ferrara, A.,
	Iliev, I.~T., Iocco, F., Jelic, V., Jensen, H., Joseph, R.,
	Labroupoulos, P., Meiksin, A., Mesinger, A., Offringa, A.~R., Pandey, V.~N.,
	Pritchard, J.~R., Santos, M.~G., Schwarz, D.~J., Semelin, B., Vedantham, H.,
	Yatawatta, S., \& Zaroubi, S.\ 2013,
	Reionization and the Cosmic Dawn with the Square Kilometre Array,
	Experimental Astronomy, 36, 235


\bibitem[Melott \etal(1989)]{Melott:1989wv}
  Melott, A.~L., Cohen, A.~P., Hamilton, A.~J.~S., Gott, J.~R., \& Weinberg, D.~H.\ 1989,
	Topology of Large Scale Structure. 4. Topology in Two-Dimensions,
  \apj, 345, 618

\bibitem[Mesinger \etal(2013)]{Mesinger:2013}
	Mesinger, A., Ferrara, A., \& Spiegel, D.~S.\ 2013,
	Signatures of X-rays in the early Universe,
	\mnras, 431, 621

\bibitem[Morales \& Hewitt(2004)]{Morales:2003vn}
  Morales, M.~F., \& Hewitt, J.\ 2004,
  Toward Epoch of Re-ionization Measurements with Wide-Field LOFAR
  Observations,
  \apj, 615, 7

\bibitem[Park \etal(1992)]{Park:1991wp}
  Park, C., Gott, J.~R., Melott, A.~L., \& Karachentsev, I.~D.\ 1992,
  The Topology of Large Scale Structure. 6. Slices of the Universe,
  \apj, 387, 1

\bibitem[Park \etal(1998)]{Park:1997ad}
  Park, C., Colley,W.~N., Gott, J.~R., Ratra, B., Spergel, D.~N., \& Sugiyama, N.\ 1998,
  CMB Anisotropy Correlation Function and Topology from Simulated Maps for
  MAP,
  \apj, 506, 473

\bibitem[Park, Gott \& Choi(2001)]{Park:2000ne}
  Park, C., Gott, J.~R., \& Choi Y.~J.\ 2001,
  Topology of the Galaxy Distribution in the Hubble Deep Fields,
  \apj, 553, 33

\bibitem[Park \etal(2001)]{Park:2001nq}
  Park, C.~G., Park, C., Ratra, B., \& Tegmark, M.\ 2001,
  Gaussianity of Degree-Scale Cosmic Microwave Background Anisotropy
  Observations,
  \apj, 556, 582

\bibitem[Park \& Kim(2010)]{Park:2009ja}
  Park, C., \& Kim, Y.~R.\ 2010,
  Large-Scale Structure of the Universe as a Cosmic Standard Ruler,
  \apjl, 715, L185

\bibitem[Parsons \etal(2010)]{Parsons:2010}
	Parsons, A.~R., Backer, D.~C., Foster, G.~S., Wright, M.~C.~H., Bradley, R.~F.,
	Gugliucci, N.~E., Parashare, C.~R., Benoit, E.~E., Aguirre, J.~E., Jacobs, D.~C.,
	Carilli, C.~L., Herne, D., Lynch, M.~J., Manley, J.~R., \& Werthimer, D.~J.\ 2010,
	The Precision Array for Probing the Epoch of Re-ionization:
	Eight Station Results,
  \aj, 139, 1468

\bibitem[Pritchard \& Loeb(2008)]{Pritchard:2008}
	Pritchard, J.~R., \& Loeb, A.\ 2008,
	Evolution of the 21cm signal throughout cosmic history,
	\prd, 78, 103511

\bibitem[Purcell \& Field(1956)]{Purcell:1956}
	Purcell, E.~M., \& Field, G.~B.\ 1956,
	Influence of Collisions upon Population of Hyperfine States in Hydrogen,
	\apj, 124, 542
	
\bibitem[Rees(1986)]{Rees:1986}
  Rees, M.~J.\ 1986,
  Lyman Absorption Lines in Quasar Spectra -- Evidence for Gravitationally-Confined Gas in Dark Minihaloes,
  \mnras, 218, 25P

\bibitem[Shapiro \etal(2006)]{Shapiro:2006}
  Shapiro, P.~R., Ahn, K., Alvarez, M., Iliev, I.~T., Martel, H., \& Ryu, D.\ 2006,
	The 21 cm Background from the Cosmic Dark Ages: Minihalos and the Intergalactic Medium before Reionization,
  \apj, 646, 681

\bibitem[Shapiro \etal(2013)]{Shapiro:2013}
	Shapiro, P.~R., Mao, Y., Iliev, I.~T., Mellema, G., Datta, K.~K.,
	Ahn, K., \& Koda, J.\ 2013,
	Will Nonlinear Peculiar Velocity and Inhomogeneous Reionization Spoil 21 cm Cosmology from the Epoch of Reionization?,
	\prl, 110, 151301

\bibitem[Shin, Trac \& Cen(2008)]{Shin:2008}
  Shin, M.-S., Trac, H., \& Cen, R.\ 2008,
  Cosmological H II Bubble Growth during Reionization,
  \apj, 681, 756

\bibitem[Smoot \etal(1992)]{Smoot:1992td}
	Smoot, G.~F., Bennett, C.~L., Kogut, A., Wright, E.~L., Aymon, J.,
	Boggess, N.~W., Cheng, E.~S., de Amici, G., Gulkis, S., Hauser, M.~G.,
	Hinshaw, G., Jackson, P.~D., Janssen, M., Kaita, E., Kelsall, T.,
	Keegstra, P., Lineweaver, C., Loewenstein, K., Lubin, P., Mather, J.,
	Meyer, S.~S., Moseley, S.~H., Murdock, T., Rokke, L., Silverberg, R.~F.,
	Tenorio, L., Weiss, R., \& Wilkinson, D.~T.\ 1992,
  Structure in the COBE Differential Microwave Radiometer First Year Maps,
  \apj, 396, L1

\bibitem[Wyithe \& Loeb(2003)]{Wyithe:2003rr}
  Wyithe, J.~S.~B., \& Loeb, A.\ 2003,
  Was the Universe Reionized by Massive Population-III Stars?,
  \apj, 588, L69

\bibitem[Zahn \etal(2006)]{Zahn:2006sg}
  Zahn, O., Lidz, A., McQuinn, M., Dutta, S., Hernquist, L., Zaldarriaga, M., \& Furlanetto, S.~R.\ 2006,
  Simulations and Analytic Calculations of Bubble Growth During Hydrogen
  Reionization,
  \apj, 654, 12

\end{thebibliography}
\end{document}